\documentclass[twocolumn]{IEEEtran}
\usepackage{amssymb}
\usepackage{}
\usepackage{amsfonts}
\usepackage{amsmath}
\usepackage{algorithm}
\usepackage{algorithmic}
\usepackage{cite}
\usepackage{subeqnarray}
\usepackage{cases}
\usepackage{stfloats}
\usepackage{graphicx,subfigure,amsmath,amssymb,cite}

\usepackage[dvips]{color}
\usepackage{float}
\ifCLASSINFOpdf
\else
\fi
\begin{document}

\title{Transmit Antenna Selection and Beamformer Design for Secure Spatial Modulation with rough CSI of Eve}

\author{Guiyang Xia,~Yan Lin,~Tingting Liu,~Feng Shu and Lajos Hanzo

%\thanks{This work was supported in part by the National Natural Science Foundation of China (Nos. 61771244, 61501238, 61702258, 61472190, and 61271230), in part by the Open Research Fund of National Key Laboratory of Electromagnetic Environment, China Research Institute of Radiowave Propagation (No. 201500013), in part by the Jiangsu Provincial Science Foundation under Project BK20150786, in part by the Specially Appointed Professor Program in Jiangsu Province, 2015, in part by the Fundamental Research Funds for the Central Universities under Grant 30916011205, and in part by the open research fund of National Mobile Communications Research Laboratory, Southeast University, China (Nos. 2017D04 and 2013D02).}
\thanks{Guiyang Xia,~Yan Lin, and Feng Shu are with the School of Electronic and Optical Engineering, Nanjing University of Science and Technology, 210094, China. Email:
xiaguiyang@njust.edu.cn;
yanlin@njust.edu.cn;
shufeng@njust.edu.cn.
}
\thanks{Tingting Liu is with the School of Information and Communication Engineering, Nanjing Institute of Technology, Nanjing 211167, China. Email:liutt@njit.edu.cn.}
%\thanks{Feng Shu is also with the School of Computer and Information at Fujian Agriculture and Forestry University, Fuzhou, 350002, China.}
\thanks{Lajos Hanzo is with the school of Electronics and Computer Science, University of Southampton, Southampton SO17 1BJ, U.K. Email:lh@ecs.soton.ac.uk.}
}
\markboth{}%
{Shell \MakeLowercase{\textit{et al.}}: Bare Demo of IEEEtran.cls for Journals}

\maketitle
\vspace{-2cm}

\begin{abstract}
The security of spatial modulation (SM) aided networks can always be improved by reducing the desired link's power at the cost of degrading its bit error ratio performance and assuming the power consumed to artificial noise (AN) projection (ANP). We formulate the joint optimization problem of maximizing the secrecy rate (Max-SR) over the transmit antenna selection and ANP in the context of secure SM-aided networks, which is mathematically a non-linear mixed integer programming problem. In order to solve this problem, we provide a pair of solutions, namely joint and separate solutions. Specifically, an accurate approximation of the SR is used for reducing the computational complexity, and the optimal AN covariance matrix (ANCM) is found by convex optimization for any given active antenna group (AAG). Then, given a large set of AAGs, simulated annealing mechanism is invoked for optimizing the choice of AAG, where the corresponding ANCM is recomputed by this optimization method as well when the AAG changes.
To further reduce the complexity of the above-mentioned joint optimization, a low-complexity two-stage separate optimization method is also proposed.
Furthermore, when the number of transmit antennas tends to infinity, the Max-SR problem becomes equivalent to that of maximizing the ratio of the desired user's signal-to-interference-plus-noise ratio to the eavesdropper's. Thus our original problem reduces to a fractional programming problem, hence a significant computational complexity reduction can be achieved for the optimization problem.
Our simulation results show that the proposed algorithms outperform the existing leakage-based null-space projection scheme in terms of the SR performance attained, and drastically reduces the complexity at a slight SR performance reduction.
\end{abstract}


\IEEEpeerreviewmaketitle

\section{Introduction}

\IEEEPARstart{A}{s} a promising technique, spatial modulation (SM) \cite{Mesleh2008Spatial} invokes the more general index modulation (IM) concept for the transmit antennas (TAs) to convey extra information \cite{Basar2017} \cite{YangP2013}, which has attracted tremendous attention over the past decade.
Recently, SM has shown advantages in terms of its spectral efficiency versus energy efficiency in various communication networks, including cooperative, full-duplex, single/multi-user and cognitive radio systems. Therefore, SM has become a popular candidate for next-generation systems \cite{Basar2016,HeL2017,Renzo2013Spatial,YangP2016Precod,Lakshmi2015,WuQ2017}.
Due to the broadcast nature of radio propagation, its security problem has to be considered, since various types of wireless access devices may overhear the private messages. Traditionally, the security of a communication system has been ensured through cryptography and authentication in the network layer, which often imposes additional computational complexity for key generation and complex decryption algorithms \cite{Jin2017} \cite{ChenX2017}. However, the key distribution and management is challenging for large-scale wireless networks. Nevertheless, physical layer security (PLS) \cite{ZouY2016} \cite{WangF2018} does not require a key for ensuring security, where the fundamental philosophy is to exploit the randomness of communication channels. Then, the transmitter (Alice) aims for conveying private information securely to the desired receiver and to keep the illegitimate receiver as ignorant of the private information as possible \cite{Wang2012Distributed,Wu2012Linear,Wu2017Secure}.

To improve the security against an eavesdropper, several SM-based PLS schemes have been proposed. In \cite{Liu2017Secure}, a full-duplex receiver was employed at the desired receiver (Bob), where Bob receives confidential messages and simultaneously transmits artificial noise (AN) to corrupt the illegal receiver (Eve). In \cite{Wang2015Secrecy} and \cite{Chen2016Secure}, AN was transmitted along the null-space of the legitimate channel for enhancing its security without any prior knowledge of Eve's location. A precoding-aided SM (PSM) scheme was proposed in \cite{WuF2016} to exploit the index of receive antennas to convey spatial information, where the precoding matrix was designed for maximizing the signal-to-noise ratio (SNR) at the desired receiver, whilst minimizing the eavesdropper's SNR.
Another interesting proposal of Wu \textit{et al.} \cite{Wu2015Secret} was that of injecting AN in the null-space of the legitimate channel to combat passive eavesdroppers. Additionally, several time-varying mapping schemes were proposed in \cite{Jiang2018} and  \cite{YangY2018} to enhance the security of SM systems, where the time-varying characteristic was the unique channel fading, which was only known to transmitter and to the legitimate receiver. Nevertheless, these researches have not considered the effect of a varying number of antennas on security.

When the number of TAs is not a power of two, selecting an active antenna group (AAG) can be adopted to further improve the security of SM systems \cite{Shu2018two} \cite{Xia2018AS}. Although the AAG selection based on the minimum Euclidean distance criterion and minimizing the bit-error-rate (BER) has drawn considerable attention in conventional SM systems \cite{Rajashekar2013Antenna,Zhou2014Reduced,Sun2017Transmit,YangP2016}, none of these preceding contributions have considered the presence of eavesdroppers.
Recently, the AAG selection scheme was also been highlighted as an efficient way of enhancing the security and creating a secure SM (SSM) network. To be specific, with the aid of AN, a leakage-based null-space projection (NSP) method (LNSP) was proposed in \cite{Shu2018two} operating at an extremely low-complexity, which achieved an acceptable secrecy rate (SR) performance. In \cite{Xia2018AS}, the authors proposed a set of AAG selection schemes for maximizing the SR. However, these AAG selection schemes only considered the SR performance gain gleaned from channel diversity, which causes a serious SR performance degradation in the high-SNR region.

As mentioned, AN was used to improve the SR performance by directly projecting into the null-space of the legitimate channel, which was achieved by exploiting its closed-form expression. However, this approach has its limitations, because it only allows the AN avoid affecting the detection of the desired receiver, but dispenses with more holistic considerations.
In addition, for circumventing that the size of the AAG combinations grows exponentially upon increasing the number of TAs, simulated annealing (SA) \cite{WangBo2018} \cite{Zhan2018} was invoked for reducing the search complexity of this combinatorial optimization problem.
The key benefit of SA is that it avoids convergence to local minima.
In this context, we assume that the rough channel state information (CSI) of the illegal channel can be obtained at the transmitter, and then exploit the PLS of an SSM system, where both TA selection and AN design are invoked for enhancing the security. The main contributions of this paper can be summarized as follows:
\begin{enumerate}
 \item To enhance the security of SM system, a joint AAG selection and AN projection design problem is formulated, which is a mixed integer programming problem.
     The original optimization problem may be solved by exhaustive search (ES) over the set of AAGs and gradient descend (GD) to compute the AN projection matrix (ANPM), in which the optimal ANPM is obtained by GD for each AAG. However, the computational complexity increases exponentially. To reduce the complexity of original SR expression involving multiple integrals, a lower bound of the approximate mutual information (AMI) between Alice and Bob and an upper bound of the AMI between Alice and Eve are derived. Then, a simpler approximate SR (ASR) is defined as the difference between the lower and upper bounds. Given a fixed AAG, maximizing the ASR (Max-ASR) problem over the AN covariance matrix (ANCM) is formulated as a convex problem.
 \item In order to solve the intractable $0$-$1$ combinatorial optimization of AAG selection, the SA mechanism is invoked for approaching the global optimum. Once a new AAG is generated, the Max-ASR criterion is applied to compute the optimal ANCM, repeatedly. This method is termed as the joint SA-Max-ASR scheme, which converges asymptotically to the globally optimal solution. Noting that the ANCM has to be designed once the AAG changes in the joint SA-Max-ASR scheme, a low-complexity two-stage algorithm referred to as the separate SA-Max-ASR is proposed, where the optimal AAG is firstly obtained based on the SA strategy, and the ANCM is designed by the Max-ASR technique, given the optimal AAG. The proposed separate SA-Max-ASR achieves a good SR performance at an extremely low complexity.
 \item As the number of TAs is increased, the two-layer summation operation over the legitimate combination of TAs and modulation symbols in the ASR expression are eliminated and a simple expression is derived. Then, maximizing the SR can be converted into maximizing the ratio of the signal-to-interference-plus-noise ratio (SINR) at the desired receiver to that at the eavesdropper. Correspondingly, a low-complexity method is presented to efficiently design the AAG and the ANCM. Our simulation results verify that the SR performance of the proposed Max-R-SINR scheme is capable of approaching the upper bound of the secrecy capacity in the high-SNR region.
\end{enumerate}

The remainder of this treatise is organized as follows. In Section II, an SSM system is described and the definition of its average SR is given. Subsequently, a GD-based method of optimizing the ANPM is proposed. In Section III, a closed-form ASR expression is derived and a concave maximization problem is formulated. Then an SA-based AAG selection scheme is presented. In Section IV, a simple optimization objective function (OF) is derived to replace the original function as the number of TAs tends to a large value, and a low complexity scheme is presented. In Section V, both the convergence and the complexity of the proposed methods is analysed, followed by our numerical results in Section VI. Finally, Section VII offers our conclusions.

\emph{Notations:} Vectors and matrices are represented in boldface.  $\left\lfloor \cdot \right\rfloor$, $\| \cdot \|^2$, $\textsf{C}_{m}^{n}$ denote the floor function, the Frobenius norm and the binomial coefficient respectively. The superscripts $\left( \cdot \right)^T$ and $\left( \cdot \right)^H$ represent the transpose and the conjugate transpose operations respectively. Besides, \textsf{tr}$(\cdot)$, \textsf{diag}$(\cdot)$ and \textsf{det}($\cdot$) denote the trace, the determinant and the diagonal of a matrix, respectively. $\mathbb{E}(\cdot)$ means the expectation operation. Matrix $\textbf{I}_N$ refers to the $N$-by-$N$ identity matrix. $\mathcal{CN}(\mu,\sigma^2)$ implies a complex Gaussian distribution with $\mu$ mean and $\sigma^2$ variance.

\section{System Model and Problem Formulation}

%Consider the SSM system illustrated in Fig.~\ref{SystemModel}, where $N_t$ transmit antennas (TAs) are employed at Alice, while both Bob and Eve are equipped with a single antenna. Here, Eve intends to intercept the confidential messages from Alice to Bob.
Consider an SSM system, where $N_t$ transmit antennas (TAs) are employed at Alice, while both Bob and Eve are equipped with a single antenna. Here, Eve intends to intercept the confidential messages from Alice to Bob.

%\begin{figure} [tp!]
% \centering
% \includegraphics[width=7cm,height=10.5cm]{SystemModel.eps}\\
% \caption{A secure SM system model.}\label{SystemModel}
%\end{figure}

Considering the fact that $N_t$ is not always a power of two, it becomes necessary to select an AAG choosing $N_s=2^{\lfloor \textrm{log}_2 N_t \rfloor}$  elements. In accordance with the concept of SM, Alice activates one out of $N_s$ antenna indices to convey $\lfloor \textrm{log}_2 N_t \rfloor$ bits of information. Then, $\lfloor \textrm{log}_2 N_t \rfloor + \textrm{log}_2M$ bits can be transmitted per channel use in total, where $M$ is the size of the $\mathcal{M}$-ary classic modulation constellation pointer.
Additionally, it can be observed that there are $L=\textsf{C}^{N_s}_{N_t}$ possible AAG combinations. Let us denote the set of all possible AAGs as $\mathcal{C}=\left\{ \mathcal{C}_1, \mathcal{C}_2,\cdots, \mathcal{C}_L \right\}$, where $\mathcal{C}_l \left(l=1, \cdots, L\right)$ denotes the $l$-th AAG and each AAG contains $N_s$ TAs.

For a given AAG $\mathcal{C}_l$, an SM symbol $\boldsymbol{x}$ associated with AN is given by
\begin{align} \nonumber
\boldsymbol{x}&=\sqrt{P_1} \textbf{s}_i^j + \sqrt{ P_2}\textbf{T}\textbf{n} \\
&=\sqrt{P_1} \textbf{e}_ib_j+ \sqrt{P_2}\textbf{T}\textbf{n},
\end{align}
where $P_1$ and $P_2$ represent the power associated with $P_1 + P_2 =P_s$, and $P_s$ denotes the total transmit power. Vector $\textbf{e}_i$ represents the $i$-th column of $\textbf{I}_{N_s}$, which implies that the $i$-th ($i=1, \cdots, N_s$) TA of $\mathcal{C}_l$ is activated. Moreover, $b_j \in \mathcal{M} = \{b_1, \cdots, b_M\}$ is the $j$-th amplitude phase modulation (APM) symbol in one $\mathcal{M}$-ary constellation. Additionally, $\textbf{n} \in \mathcal{CN}(\textbf{0}, \textbf{I}_{N_s})$ is the AN vector, and $\textbf{T}$ is the ANPM with $\mathbb{E} \left(\textbf{T}\textbf{T}^H \right) =\textbf{Q}$, where $\textbf{Q}$ is the ANCM and satisfy $\textsf{tr}(\textbf{Q})=1$. The corresponding signals received at Bob and at Eve are respectively represented as
\begin{align} \nonumber
y_B &=\textbf{h} \textbf{S}_l \boldsymbol{x}+ n_B \\ \label{y_B}
&=\sqrt{P_1}\textbf{h}_l\textbf{e}_ib_j + \sqrt{ P_2}\textbf{h}_l\textbf{T}\textbf{n}+ n_B, \\ \nonumber
y_E &=\textbf{g} \textbf{S}_l \boldsymbol{x} + n_E \\ \label{y_E}
&=\sqrt{P_1}\textbf{g}_l\textbf{e}_ib_j +\sqrt{ P_2} \textbf{g}_l \textbf{T}\textbf{n}+ n_E,
\end{align}
where $\textbf{S}_l$ is the AAG selection matrix, while $\textbf{h}_{l}= \textbf{h} \textbf{S}_l \in \mathbb{C} ^{1 \times N_s}$ and $\textbf{g}_{l}= \textbf{g} \textbf{S}_l \in \mathbb{C} ^{1 \times N_s}$ are the sub-channels of $\textbf{h}$ and $\textbf{g}$ that depends on $\mathcal{C}_l$. Furthermore, $\textbf{h} \in \mathbb{C}^ {1 \times N_t}$ and $\textbf{g} \in \mathbb{C}^ {1 \times N_t}$ are the complex channel gain vectors spanning from Alice to Bob and from Alice to Eve, respectively. Additionally, $n_B \in \mathcal{CN}(0, \sigma_B^2)$ and $n_E \in \mathcal{CN}(0, \sigma_E^2)$ are the independent complex Gaussian noises at Bob and at Eve. Based upon the received signal in (\ref{y_B}) for a given $\mathcal{C}_l$, the maximum likelihood (ML) detector may be utilized by Bob to jointly detect the spatial symbol and the conventional APM symbol, formulated as:
\begin{align} \label{ML_detection}
( \hat{i}, \hat{j} )= \arg \min \limits_\mathcal{S}  | y_B-\sqrt{P_1}\textbf{h}_{l}\textbf{e}_ib_j |^2,
\end{align}
where $\mathcal{S}=\mathcal{C}_l \times \mathcal{M}$ is the super-alphabet set that contains all possible combinations of the active antennas and the conventional symbols.

In this context, we assume that Alice has perfect CSI on the main channel. An accurate CSI estimate of the channel from Alice to Bob may be derived by using training sequences, which is then sent back to the transmitter through dedicated feedback links \cite{Wu2017Secure}. Additionally, it is also assumed that Alice can obtain a rough CSI estimate of the eavesdropper's channel, which corresponds to the scenario that Eve is an active user in wireless networks. According to \cite{Tang2013} \cite{Yu2018}, the additive uncertainly model for the CSI of Eve at Alice is given by
\begin{align}
\textbf{g}= \tilde {\textbf{g}} + \boldsymbol{\Delta}{\textbf{g}},
\end{align}
where $\tilde {\textbf{g}}$ is the estimated channel of $\textbf{g}$, and $\boldsymbol{\Delta} {\textbf{g}}$ is the corresponding estimation error, which is actually a zero-mean Gaussian random variable associated with the covariance $\sigma_e^2$, i.e.,  $\boldsymbol{\Delta} {\textbf{g}} \sim \mathcal{CN} (\textbf{0}, \sigma_e^2 \textbf{I})$. In the remainder of this treatise, we will develop schemes of enhancing the security of such SSM networks.

In general, the transmit symbol $b_j$ is equiprobably drawn from a discrete $\mathcal{M}$-ary constellation. For a specific channel realization and fixed $\mathcal{C}_l$, the mutual information (MI) between Alice and Bob can be expressed as
\begin{align} \nonumber
I(& \boldsymbol{x} ;y_B)=\textrm{log}_2N_sM - \frac{1}{N_sM} \times \\  \label{I_B}
&~ \sum \limits_{i  = 1}^{N_sM} \mathbb{E}_{n_B} \left\{ \textrm{log}_2 \sum \limits_{j=1} ^{N_sM} \textrm{exp} \left( \frac{-f_{b,i,j} }{ P_2 \textbf{h}_l \textbf{Q} \textbf{h}_l^H + \sigma_B^2} \right) \right\},
\end{align}
where
\begin{align}
f_{b,i,j}&=|\sqrt{P_1} \textbf{h}_l(\boldsymbol{x}_i-
\boldsymbol{x}_j)+ n_B|^2 -|n_B|^2,
\end{align}
and $\boldsymbol{x}_{\left\{\cdot\right\}}$ represents a legitimate transmit symbol in the set $\mathcal{S}$. Similarly, we have the MI between Alice and Eve:
\begin{align}  \nonumber
I(&\boldsymbol{x};y_E)=\textrm{log}_2N_sM - \frac{1}{N_sM} \sum \limits_{m  = 1}^{N_sM} \mathbb{E}_{\boldsymbol{\Delta} {\textbf{g}},n_E} \\ \label{I_E}
&\left\{ \textrm{log}_2 \sum \limits_{k=1} ^{N_sM} {\textrm{exp} \left( \frac{-f_{e,m,k} }{ P_2 (\tilde {\textbf{g}}_l  + \boldsymbol{\Delta}{\textbf{g}}_l) \textbf{Q} (\tilde {\textbf{g}}_l  + \boldsymbol{\Delta}{\textbf{g}}_l)^H + \sigma_E^2} \right)} \right\},
\end{align}
where
\begin{align}
f_{e,m,k}&=|\sqrt{P_1} (\tilde {\textbf{g}}_l  + \boldsymbol{\Delta}{\textbf{g}}_l)(\boldsymbol{x}_m-\boldsymbol{x}_k)+n_E|^2 -|n_E|^2.
\end{align}
Combining (\ref{I_B}) and  (\ref{I_E}) yields the ergodic SR  defined as
\begin{align} \label{Average_SR}
\bar{R}_s =\mathbb{E}_{\textbf{h},\textbf{g}}\left[ I(\boldsymbol{x};y_B)- I(\boldsymbol{x};y_E),0 \right]^+,
\end{align}
where $\left[ a \right]^+$=max(a,~0) and $R_s =I(\boldsymbol{x};y_B)- I(\boldsymbol{x};y_E)$ is the instantaneous SR for a specific channel realization. Since the number of TAs is not a power of two, it is necessary to select an AAG and to design AN for enhancing the security. From the definition of (\ref{Average_SR}), the optimization problem of maximizing the SR (Max-SR) can be cast as
\begin{subequations}  \label{P0}
\begin{align}
&\max\limits_{ \textbf{s}_l, \textbf{Q}} ~~ R_s \label{Target_O}\\
&s.t.~~ \textsf{tr}(\textbf{Q}) =1, \label{Constraint_Q} \\
&~~~~~~ \textbf{Q} \succeq 0, \label{Power} \\
&~~~~~~ \sum \limits_{i=1}^{N_t} s_i =N_s, \label{TASelection} \\
&~~~~~~ s_i \in \left\{ 0,1 \right\}, i=1, \cdots, N_t, \label{IntegerConstraint}
\end{align}
\end{subequations}
where $s_i$ is the $i$-th element of the AAG vector $\textbf{s} \in R^{N_t \times 1}$. Mathematically, $N_s$ '1' elements are assigned to the diagonal locations of a $N_t\times N_t$ matrix of zeros. For instance, we have to select $4$ TAs to form an AAG when $N_t=6$. If the $5$-th AAG of say $\textbf{s}_5=\left\{1,2,4,6\right\}$, has the optimal SR performance, the corresponding AAG selection matrix becomes:
\begin{align}
\textbf{S}_{5} = \textsf{diag} (\textbf{s}_5)  \buildrel \Delta \over =  \left( \begin{array}{cccccc}
         1 & 0 & 0 & 0 & 0 & 0\\
         0 & 1 & 0 & 0 & 0 & 0\\
         0 & 0 & 0 & 1 & 0 & 0\\
         0 & 0 & 0 & 0 & 0 & 1
       \end{array}
 \right) ^T.
\end{align}
For a given $\textbf{S}_{l}$, the actual effective sub-channel spanning from Alice to Bob is $\textbf{h}_l=\textbf{h}\textbf{S}_l$. Similarly, the sub-channel from Alice to Eve is $\textbf{g}_l=\textbf{g}\textbf{S}_l$. Observe from (\ref{P0}) that the optimization problem is a mixed integer programming problem, where (\ref{Constraint_Q}) and (\ref{Power}) are continuous constraints whereas (\ref{TASelection}) and (\ref{IntegerConstraint}) are 0-1 integer constraints.
More particularly, due to the constraints (\ref{TASelection}) and (\ref{IntegerConstraint}), the optimization problem in (\ref{P0}) is an NP-hard problem, hence it requires an excessive search complexity to find the optimal subset of maximizing $R_s$. Additionally, the OF is non-convex and it is not in closed-form, thus it is a challenging task to solve problem (\ref{P0}).

\section{AAG selection and AN design for maximizing the SR performance}

In this section, a GD-based method is first presented to design the ANPM, where ES is adopted to find the optimal AAG. This scheme is used as our benchmark. Then, a joint SA-Max-ASR scheme of AAG and ANCM is proposed to achieve a high SR performance. However, to reduce the complexity, a low-complexity separate SA-Max-ASR scheme is also proposed at the cost of a slight performance loss.

\subsection{GD-based design of AN projection matrix}

Due to the non-convexity of the OF $R_s$, obtaining a closed-form solution for (\ref{P0}) becomes intractable. However, it is natural to adopt numerical algorithms to search for local maxima of the OF such as GD.
Given a fixed $\textbf{S}_l$, $\textbf{T}$ can be optimized first, and then averaged over all possible AAGs to find the best $\textbf{S}_l$ and $\textbf{T}$. The optimization problem can be rewritten as
\begin{subequations}  \label{P_Direct}
\begin{align}
&\max\limits_{\textbf{T}} ~~ R_s \left( \textbf{S}_l \right) \\
&s.t.~~ \textsf{tr}(\textbf{T}\textbf{T}^H) =1.
\end{align}
\end{subequations}
Accordingly, the Lagrangian function can be directly written as:
\begin{align}
L\left( \textbf{T}, \lambda \right)= -R_s \left( \textbf{S}_l\right) + \lambda \left[ \textsf{tr} (\textbf{T}\textbf{T}^H) -1 \right],
\end{align}
where $\lambda$ is the Lagrange multiplier. Then, the GD-based method can be applied to solve this optimization problem and its detailed process is illustrated in Algorithm \ref{Alg1}.
By taking the gradient of $R_s \left(\textbf{S}_l\right)$ with respect to $\textbf{T}$  and setting it to zero, we have
\begin{align}
-\nabla_\textbf{T} R_s \left( \textbf{S}_l\right)  +\lambda \textbf{T} =0,
\end{align}
where $\nabla_\textbf{T} R_s \left(\textbf{S}_l\right)$ is shown in (\ref{Grad_Rs}), and

\begin{figure*}
\begin{align} \nonumber
& \nabla_\textbf{T} R_s(\textbf{S}_l) =  \frac{1}{\textrm{ln}2 \cdot N_sM} \times  \\ \nonumber
& \left\{ \sum \limits_{m  = 1}^{N_sM} \mathbb{E}_{\boldsymbol{\Delta} {\textbf{g}_l},n_E}\left( \frac{1}{ \kappa_E} \sum \limits_{k=1} ^{N_sM} \left( \frac{ P_2 f_{e,m,k} (\tilde {\textbf{g}}_l  + \boldsymbol{\Delta}{\textbf{g}}_l)^H(\tilde {\textbf{g}}_l  + \boldsymbol{\Delta}{\textbf{g}}_l) \textbf{T} }{ \left( P_2(\tilde {\textbf{g}}_l  + \boldsymbol{\Delta}{\textbf{g}}_l) \textbf{T}\textbf{T}^H (\tilde {\textbf{g}}_l  + \boldsymbol{\Delta}{\textbf{g}}_l)^H + \sigma_E^2\right)^2 } \right) {\textrm{exp} \left( \frac{-f_{e,m,k} }{ P_2 (\tilde {\textbf{g}}_l  + \boldsymbol{\Delta}{\textbf{g}}_l) \textbf{T}\textbf{T}^H (\tilde {\textbf{g}}_l  + \boldsymbol{\Delta}{\textbf{g}}_l)^H + \sigma_E^2} \right)}\right) \right. \\
& ~~~~~ -\left. \sum \limits_{i  = 1}^{N_sM} {\mathbb{E}_{n_B} \left( \frac{1}{ \kappa_B} \sum \limits_{j=1} ^{N_sM} \left( \frac{P_2 f_{b,i,j} \textbf{h}_l^H\textbf{h}_l\textbf{T} }{\left( P_2 \textbf{h}_l \textbf{T}\textbf{T}^H \textbf{h}_l^H + \sigma_B^2\right)^2 } \right) {\textrm{exp} \left( \frac{-f_{b,i,j} }{ P_2 \textbf{h}_l \textbf{T}\textbf{T}^H \textbf{h}_l^H + \sigma_B^2} \right)}\right)}   \right\}   \label{Grad_Rs}
\end{align}
\hrulefill
\end{figure*}

\begin{align} \label{Grad_Sub1}
&\kappa_E = \sum \limits_{k=1} ^{N_sM} {\textrm{exp} \left( \frac{-f_{e,m,k} }{ P_2 (\tilde {\textbf{g}}_l  + \boldsymbol{\Delta}{\textbf{g}}_l) \textbf{T}\textbf{T}^H (\tilde {\textbf{g}}_l  + \boldsymbol{\Delta}{\textbf{g}}_l)^H + \sigma_E^2} \right)}, \\  \label{Grad_Sub2}
&\kappa_B = \sum \limits_{j=1} ^{N_sM} {\textrm{exp} \left( \frac{-f_{b,i,j} }{ P_2 \textbf{h}_l \textbf{T}\textbf{T}^H \textbf{h}_l^H + \sigma_B^2} \right)}.
\end{align}
Next, by substituting (\ref{Grad_Sub1}) and (\ref{Grad_Sub2}) into (\ref{Grad_Rs}), the ES plus GD method shown in Algorithm \ref{Alg1} can be exploited to obtain $\textbf{S}_l$ and $\textbf{T}$ numerically. Hence, Algorithm \ref{Alg1} can be guaranteed to converge to a local optimum. Finally, the optimal $\textbf{S}^\ast$ and $\textbf{T}^\ast$ may be obtained by repeating Algorithm \ref{Alg1} using a number of distinct initializations.
\begin{algorithm}[tp!]
\caption{Numerical search for maximizing SR (ES plus GD)} \label{Alg1}
\begin{algorithmic}[1]
\STATE List all possible $\textbf{S}_l$, $l = (1, \cdots, L)$.
\STATE \textbf{For} $ l = 1: L $
\STATE Initial $\textbf{T}_1$ with constraint $\textsf{tr}(\textbf{T}_1\textbf{T}_1^H)=1$. Set step size $\mu$ and minimum tolerance $\mu_{min}$.
\STATE Set $k=1$, calculate $R_s(k)=R_s(\textbf{S}_l,\textbf{T}_1)$.
\STATE If $\mu \geq \mu_{min}$ goto step $6$, otherwise stop algorithm and return $\textbf{T}_k$.
\STATE Calculate $\textbf{T}_k' = \textbf{T}_k + \mu \nabla_{\textbf{T}_k} R_s(\textbf{S}_l)$, and normalize $\textbf{T}_k'$.
\STATE Calculate $R_s' = R_s(\textbf{T}_k')$.
\STATE If $R_s' \geq R_s(\textbf{T}_k)$, update $R_s(k+1) = R_s'$ and $\textbf{T}_{k+1} = \textbf{T}_k'$, then goto step $9$; Otherwise, $\mu=\mu/2$ and goto step $5$.
\STATE $k = k+1$ goto step 6.
\STATE Storing $\textbf{R}(l) =[R_{s,k+1}, \textbf{T}_{k+1}]$.
\STATE \textbf{End}
\STATE Output $[\textbf{S}^\ast, \textbf{T}^\ast]$ = $ \arg \max \limits_l \textbf{R}(l)$.
\end{algorithmic}
\end{algorithm}

\subsection{Proposed joint SA-Max-ASR optimization of AAG and ANCM}

In the preceding subsection, the variables $\textbf{S}$ and $\textbf{T}$ were obtained by numerical search algorithm, which involves a large number of SR evaluations and combinations, when $N_t$ is large. Hence, it only can be used as a performance benchmark for small-scale scenarios. In what follows, we will present a lower complexity algorithm for optimizing $\textbf{S}$ and $\textbf{Q}=\mathbb{E}(\textbf{T}\textbf{T}^H)$, which  has the capability to approach the SR performance of Algorithm $\ref{Alg1}$.
Once the corresponding optimal solution is obtained, the ergodic SR is evaluated by (\ref{Average_SR}). We circumvent the difficulty by avoiding the calculation of multiple integrals in the SR expression, where the corresponding approximate MI (AMI) is given by
\begin{align}\label{Appro_B}
I_B^{a}=\zeta - \textrm{log}_2\sum \limits_{i=1}^{N_sM}\sum \limits_{j=1}^{N_sM}\textrm{exp} \left( \frac{- P_1 \textbf{d}_{ij}^H\textbf{h}_l^H  \textbf{h}_l\textbf{d}_{ij}}{4 \left( P_2 \textbf{h}_l\textbf{Q} \textbf{h}_l^H + \sigma_B^2 \right) } \right),
\end{align}
where $\zeta= 2\textrm{log}_2 N_sM$ and $\textbf{d}_{ij}=\boldsymbol{x}_i-\boldsymbol{x}_j$.  Similarly, the AMI for Eve is given by
\begin{align}\nonumber
I_E^{a}& = \zeta - \mathbb{E}_{\Delta \textbf{g}_l} \textrm{log}_2\sum \limits_{m=1}^{N_sM}\sum \limits_{k=1}^{N_sM} \\
& \textrm{exp}\left( \frac{- P_1 \textbf{d}_{mk}^H \left( \tilde {\textbf{g}}_l + \boldsymbol{\Delta}{\textbf{g}}_l\right)^H  \left(\tilde {\textbf{g}}_l + \boldsymbol{\Delta}{\textbf{g}}_l\right) \textbf{d}_{mk}}{4 \left( P_2 \left( \tilde {\textbf{g}}_l + \boldsymbol{\Delta}{\textbf{g}_l}\right)\textbf{Q} \left( \tilde {\textbf{g}}_l + \boldsymbol{\Delta}{\textbf{g}_l}\right)^H + \sigma_E^2 \right) } \right),  \label{Appro_E}
\end{align}
where $\textbf{d}_{mk}=\boldsymbol{x}_m-\boldsymbol{x}_k$. For a similar derivation process  please refer to the Appendix A of \cite{Aghdam2017Joint}. Then, upon replacing $R_s$ by (\ref{Appro_B}) and (\ref{Appro_E}), the ASR with a given AAG becomes
\begin{align}
R_s^a=  E(\textbf{S}_l, \textbf{Q})- B(\textbf{S}_l, \textbf{Q}),
\end{align}
where
\begin{align} \nonumber
& E(\textbf{S}_l, \textbf{Q})=   \mathbb{E}_{\Delta \textbf{g}_l} \textrm{log}_2\sum \limits_{m=1}^{N_sM}\sum \limits_{k=1}^{N_sM} \\ \label{E_Appro}
& ~~\textrm{exp}\left( \frac{- P_1 \textbf{d}_{mk}^H \left( \tilde {\textbf{g}}_l + \boldsymbol{\Delta}{\textbf{g}}_l\right)^H  \left(\tilde {\textbf{g}}_l + \boldsymbol{\Delta}{\textbf{g}}_l\right) \textbf{d}_{mk}}{4 \left( P_2 \left( \tilde {\textbf{g}}_l + \boldsymbol{\Delta}{\textbf{g}_l}\right)\textbf{Q} \left( \tilde {\textbf{g}}_l + \boldsymbol{\Delta}{\textbf{g}_l}\right)^H + \sigma_E^2 \right) } \right), \\  \nonumber
& B(\textbf{S}_l, \textbf{Q}) = \\
& ~~~~~~~\textrm{log}_2\sum \limits_{i=1}^{N_sM}\sum \limits_{j=1}^{N_sM}\textrm{exp} \left( \frac{- P_1 \textbf{d}_{ij}^H\textbf{h}_l^H  \textbf{h}_l\textbf{d}_{ij}}{4 \left( P_2 \textbf{h}_l\textbf{Q} \textbf{h}_l^H + \sigma_B^2 \right) } \right). \label{B_Appro}
\end{align}
It is noteworthy that replacing (\ref{I_B}) by the AMI (\ref{Appro_B}) is an efficient way of reducing the computational complexity \cite{Aghdam2017Joint}. Via applying Jensen's inequality,  (\ref{E_Appro}) can be lower bounded as
\begin{align} \nonumber
&\mathbb{E}_{\Delta \textbf{g}}  \textrm{log}_2\sum \limits_{m=1}^{N_sM}\sum \limits_{k=1}^{N_sM}\textrm{exp} \\ \nonumber
&~~\left( \frac{- P_1 \textbf{d}_{mk}^H \left( \tilde {\textbf{g}}_l + \boldsymbol{\Delta}{\textbf{g}_l}\right)^H \left( \tilde {\textbf{g}}_l + \boldsymbol{\Delta}{\textbf{g}_l}\right) \textbf{d}_{mk}}{4 \left( P_2 \left( \tilde {\textbf{g}}_l + \boldsymbol{\Delta}{\textbf{g}_l}\right)\textbf{Q} \left( \tilde {\textbf{g}}_l + \boldsymbol{\Delta}{\textbf{g}_l}\right)^H + \sigma_E^2 \right) } \right)  \geq  \\  \nonumber
& \textrm{log}_2\sum \limits_{m=1}^{N_sM}\sum \limits_{k=1}^{N_sM}\textrm{exp} \\
& ~~ \mathbb{E}_{\Delta \textbf{g}} \left( \frac{- P_1 \textbf{d}_{mk}^H\left( \tilde {\textbf{g}}_l + \boldsymbol{\Delta}{\textbf{g}_l}\right)^H  \left( \tilde {\textbf{g}}_l + \boldsymbol{\Delta}{\textbf{g}_l}\right)\textbf{d}_{mk}}{4 \left( P_2 \left( \tilde {\textbf{g}}_l + \boldsymbol{\Delta}{\textbf{g}_l}\right)\textbf{Q} \left( \tilde {\textbf{g}}_l + \boldsymbol{\Delta}{\textbf{g}_l}\right)^H + \sigma_E^2 \right) }\right).
\end{align} %\cite{Lin2013Appro}
According to \cite{Zhu2017}, we have
\begin{align}  \nonumber
\mathbb{E}_{\Delta \textbf{g}}& \left( \frac{- P_1 \textbf{d}_{mk}^H \left( \tilde {\textbf{g}}_l + \boldsymbol{\Delta}{\textbf{g}_l}\right)^H  \left( \tilde {\textbf{g}}_l + \boldsymbol{\Delta}{\textbf{g}_l}\right)\textbf{d}_{mk}}{4 \left( P_2 \left( \tilde {\textbf{g}}_l + \boldsymbol{\Delta}{\textbf{g}_l}\right)\textbf{Q} \left( \tilde {\textbf{g}}_l + \boldsymbol{\Delta}{\textbf{g}_l}\right)^H + \sigma_E^2 \right) } \right)  \approx \\
& \frac{- P_1 \mathbb{E}_{\Delta \textbf{g}} \left( \textbf{d}_{mk}^H\left( \tilde {\textbf{g}}_l + \boldsymbol{\Delta}{\textbf{g}_l}\right)^H  \left( \tilde {\textbf{g}}_l + \boldsymbol{\Delta}{\textbf{g}_l}\right)\textbf{d}_{mk} \right) }{4 \left( P_2 \mathbb{E}_{\Delta \textbf{g}} \left( \left( \tilde {\textbf{g}}_l + \boldsymbol{\Delta}{\textbf{g}_l}\right)\textbf{Q}\left( \tilde {\textbf{g}}_l + \boldsymbol{\Delta}{\textbf{g}_l}\right)^H \right) + \sigma_E^2 \right)}. \label{ExpectError}
\end{align}
Then, $E(\textbf{Q})$ can be rewritten as
\begin{align}  \nonumber
& \tilde{E}(\textbf{S}_l, \textbf{Q})= \\
&\textrm{log}_2\sum \limits_{m=1}^{N_sM}\sum \limits_{k=1}^{N_sM}\textrm{exp} \left\{ \frac{- P_1 \left( \textbf{d}_{mk}^H ( \tilde{\textbf{g}}_l^H  \tilde{\textbf{g}}_l +\sigma_e^2 \textbf{I} ) \textbf{d}_{mk} \right) }{4 \left( P_2  \tilde{\textbf{g}}_l \textbf{Q} \tilde{\textbf{g}}_l^H + \sigma_e^2P_2 + \sigma_E^2 \right) } \right\}.
\end{align}
Next, let us define a closed-form expression for the approximate SR (ASR) as follows
\begin{align} \label{Close_ASR}
R_A (\textbf{S}_l, \textbf{Q})= \tilde{E}(\textbf{S}_l, \textbf{Q}) - B(\textbf{S}_l, \textbf{Q}).
\end{align}
Therefore, the optimization problem in (\ref{P0}) can be converted into
\begin{subequations}  \label{P_Closed}
\begin{align} \label{Appro_SR}
&\max\limits_{\textbf{S}_l, \textbf{Q}} ~~ R_A (\textbf{S}_l, \textbf{Q}) \\
&s.t.~~ \textsf{tr}(\textbf{Q}) =1, \textbf{Q} \succeq 0, \\
&~~~~~~ \sum \limits_{i=1}^{N_t} s_i =N_s,  \\
&~~~~~~ s_i \in \left\{ 0,1 \right\}, i=1, \cdots, N_t.
\end{align}
\end{subequations}
However, (\ref{Appro_SR}) is still a non-concave function of the continuous optimization variable $\textbf{Q}$ with a fixed $\textbf{S}_l$. Next we convert $R_A (\textbf{S}_l, \textbf{Q})$ into a concave function, so that a unique solution $\textbf{Q}$ can be obtained.

To elaborate, we first derive a convex function as an upper bound of $B(\textbf{S}_l, \textbf{Q})$.
Due to the fact that the concave function of
\begin{align}
B_{ij}(\textbf{S}_l, \textbf{Q})=\frac{ - A_{ij}}{\textbf{h}_l\textbf{Q} \textbf{h}_l^H + b}
\end{align}
can be upper bounded by its first-order approximation $B_{ij}^{(1)}(\textbf{S}_l, \textbf{Q})$ , i.e., its tangent at point $\textbf{Q}_0$, thus we have
\begin{align}  \nonumber
&B_{ij}(\textbf{S}_l, \textbf{Q})  \leq B_{ij}^{(1)}(\textbf{S}_l, \textbf{Q}) \\
&~~~~~ \buildrel \Delta \over =  \frac{- A_{ij}}{ \textbf{h}_l\textbf{Q}_0 \textbf{h}_l^H + b } + \textsf{tr} \left\{ \frac{ A_{ij}\textbf{h}_l^H\textbf{h}_l }{ \left( \textbf{h}_l\textbf{Q}_0 \textbf{h}_l^H + b \right)^2 } \left( \textbf{Q}- \textbf{Q}_0 \right) \right\},  \label{Linear_B_1}
\end{align}
where $A_{ij}=P_1\textbf{d}_{ij}^H\textbf{h}_l^H \textbf{h}_l\textbf{d}_{ij}/4P_2$
and $b= \sigma_B^2/P_2$. The inequality (\ref{Linear_B_1}) holds due to
%is concave ($\textbf{Q} \succeq \textbf{0}$) with
$A_{ij} \geq0$ and $b>0$. Substituting the above inequality into (\ref{B_Appro}), we have the following inequality
\begin{align} \label{Covex_Upp}
B(\textbf{S}_l, \textbf{Q}) \leq \tilde{B}(\textbf{S}_l, \textbf{Q})= \textrm{log}_2\sum \limits_{i=1}^{N_sM}\sum \limits_{j=1}^{N_sM}\textrm{exp} \left( B_{ij}^{(1)}(\textbf{S}_l, \textbf{Q}) \right),
\end{align}
where $B_{ij}^{(1)}(\textbf{S}_l, \textbf{Q})$ is a linear function of $\textbf{Q}$. It is plausible that $\tilde B(\textbf{S}_l, \textbf{Q})$ is convex and it is also an upper bound of $B(\textbf{S}_l, \textbf{Q})$.
At the same time, considering that $B(\textbf{S}_l, \textbf{Q})$ is always larger than or equal to $0$, $B(\textbf{S}_l, \textbf{Q})$ can be further upper bounded as
\begin{align}
B(\textbf{S}_l, \textbf{Q}) \leq   B'(\textbf{S}_l, \textbf{Q})= \max \left\{ \tilde B(\textbf{S}_l, \textbf{Q}), ~0 \right\}.
\end{align}

In the following, a concave function related to the lower bound of $E(\textbf{Q})$ is also derived. Firstly, we reformulate the exponential function at a feasible point $\textbf{Q}_0$, given by
\begin{align} \nonumber
&\textrm{exp} \left( \frac{- C_{mk} }{ \tilde{\textbf{g}}_l \textbf{Q} \tilde{\textbf{g}}_l^H  + c  } \right)  \geq E_{mk}^{(1)}(\textbf{S}_l, \textbf{Q})\\
& \buildrel \Delta \over =  \textrm{exp} \left( \frac{- C_{mk} }{ \tilde{\textbf{g}}_l \textbf{Q}_0 \tilde{\textbf{g}}_l^H  + c  } \right) \cdot  \left( 1 + \frac{ C_{mk} }{ \tilde{\textbf{g}}_l \textbf{Q}_0 \tilde{\textbf{g}}_l^H  + c} -\frac{ C_{mk} }{ \tilde{\textbf{g}}_l \textbf{Q} \tilde{\textbf{g}}_l^H  + c} \right)  \label{Liear_E}
\end{align}
where
\begin{align}
C_{mk}= \frac{P_1\textbf{d}_{mk}^H \left( \tilde{\textbf{g}}_l^H  \tilde{\textbf{g}}_l +\sigma_e^2 \textbf{I} \right)\textbf{d}_{mk}}{4P_2},
\end{align}
and $c=\sigma_e^2 + \sigma_E^2/ P_2$. Explicitly, $E_{mk}^{(1)}(\textbf{S}_l, \textbf{Q})$ is a concave function of $\textbf{Q}$. Note that $E_{mk}^{(1)}(\textbf{S}_l, \textbf{Q})$ can be negative, hence we extend the domain of $\textrm{log}_2(x)$ to the field of real numbers, given by
\begin{align}
\dot{\textrm{log}}_2(x)=\left\{ \begin{array}{l}
\textrm{log}_2(x) ,~x >0 \\
~-\infty ~,~x \leq 0
\end{array} \right..
\end{align}

Using the inequality (\ref{Liear_E}), we rewrite $\tilde E(\textbf{S}_l, \textbf{Q})$ with the aid of a lower bound as follows
\begin{align} \label{Concave_Low}
&\tilde E(\textbf{S}_l, \textbf{Q}) \geq \bar E(\textbf{S}_l, \textbf{Q})= \dot{\textrm{log}}_2\sum \limits_{m=1}^{N_sM}\sum \limits_{k=1}^{N_sM} E_{mk}^{(1)}(\textbf{Q}).
\end{align}
Meanwhile, upon considering that $\bar E(\textbf{Q})$ is less than $2\textrm{log}_2N_sM$ and taking into account (\ref{Concave_Low}), we have
\begin{align}
\tilde E(\textbf{S}_l, \textbf{Q}) \geq  E'(\textbf{S}_l, \textbf{Q})= \min \left\{\bar E(\textbf{S}_l, \textbf{Q}),~ 2\textrm{log}_2N_sM \right\},
\end{align}
where $E'(\textbf{S}_l, \textbf{Q})$ is a point-wise maximum of a concave function and a constant, and thus it is also a concave function of $\textbf{Q}$ for a given $\textbf{S}_l$. By replacing $E'(\textbf{S}_l, \textbf{Q})$ and $B'(\textbf{S}_l, \textbf{Q})$ with $\tilde E(\textbf{S}_l, \textbf{Q})$ and $B(\textbf{S}_l, \textbf{Q})$ respectively, a concave maximization problem can be formulated for $\textbf{Q}$ as follows:
\begin{subequations} \label{Regular_Q}
\begin{align} \label{Regular_Expression}
\max\limits_{\textbf{S}_l,\textbf{Q}} ~~& R_s^c (\textbf{S}_l, \textbf{Q}) = E'(\textbf{S}_l, \textbf{Q}) - B'(\textbf{S}_l, \textbf{Q}) \\
s.t. ~~&\textsf{tr}(\textbf{Q})=1, \\
&\textbf{Q} \succeq 0.
\end{align}
\end{subequations}
Then, the optimal $\textbf{Q}$ for (\ref{Regular_Q}) can be obtained iteratively with a random given feasible point.

In the above section, we converted the continuous optimization problem into a concave one, thus a unique solution $\textbf{Q}$ can be obtained for any given $\textbf{S}_l$ $(l=1, \cdots, L)$.
Next we focus our attention on solving the $0$-$1$ programming problem for AAG selection. As it is widely known, in contrast to GD, SA explores the entire search space in a random guided fashion by sometimes degrading the OF value in an attempt to avoid getting trapped in local minima. In other words, the SA method iterates by perturbing the current configuration and measuring the change in cost.
When the change in cost is positive, the new AAG is automatically accepted, otherwise the probability of accepting the OF reduction is calculated by evaluating the so-called Boltzmann factor $C_k$. If this probability is higher than a random number in the interval $[0,1)$, the new AAG is accepted, otherwise, it is rejected. This acceptance criterion can be expressed as \cite{WangBo2018}
\begin{align}
\textrm{min} \left\{1, \textrm{exp} \left( \frac{ R_s^c (\textbf{S}_l, \textbf{Q})-R_s^c (\textbf{S}_l^o, \textbf{Q}^o)}{C_0} \right) \right\} > \eta,
\end{align}
where $\textbf{S}_l^o$ is a random neighbour AAG of the current $\textbf{S}_l$ and $\textbf{Q}^o$ is the corresponding ANCM computed by (\ref{Regular_Q}), while $C_0$ is the initial control parameter associated with $C_0>0$ and $\eta \in [0,1)$.

The proposed SA-based AAG selection scheme includes the following procedures: 1) generate neighbour AAG to impose a perturbation, 2) stochastic motion to avoid getting trapped in local maxima, 3) reduce the mutation parameter $C_k$ to increase the search precision. Let us denote the set of optimal points as
\begin{align}
S_{\textrm{opt}}= \left\{ \textbf{S}_l=\textsf{diag} \left(\textbf{s}_{\textrm{opt}}\right) : R_s^c (\textbf{s}_{\textrm{opt}}) \geq R_s^c (\textbf{s}), \textbf{s} \in \mathcal{C} \right\},
\end{align}
and $N(\textbf{s}) \in \mathcal{C}$ is the neighborhood of the solution $\textbf{s}$. Together with (\ref{Regular_Q}), the proposed joint SA-Max-ASR scheme using SA is listed in Algorithm $\ref{Alg2}$, where $R_s^c(\textbf{s})$ denotes $R_s^c (\textbf{S}_l, \textbf{Q})$ for simplicity.

%\makeatletter\def\@captype{table}\makeatother
\begin{algorithm} [tp!]
\caption{Joint AAG selection and AN design for maximizing ASR (Joint SA-Max-ASR)} \label{Alg2}
\begin{algorithmic} [1]
\STATE Given an initial AAG vector $\textbf{s} \in \mathcal{C}$; \\
       Initialization of simulation mutation parameters $C_0 > C_f >0$; \\
       Set an iterative counter $k=0$ and give a sampling number $S$.
\STATE Mutation process: \\
       1). Generating a random solution $\textbf{s}' \in N(\textbf{s})$, and evaluating $\triangle R = R_s^c(\textbf{s}') - R_s^c(\textbf{s})$. \\
       2). If 'Metropolis criterion' is satisfied, i.e., $\textrm{min}$ $\left\{ 1, \textrm{exp}(\triangle R / C_k) \right\} > \eta \in [0,1)$, then $\textbf{s}=\textbf{s}'$. \\
       3). If 'Metropolis equilibrium' under $C_k$ is realized, then go to $3$; Otherwise, go to Step $2.1$.
\STATE Integer sampling process: \\
       1). Evaluate $R_s^c(\textbf{s})$.  \\
       2). Give a temporary set $V={\textbf{s}}$, and set $R_{pre}^c = R_s^c(\textbf{s})$. \\
       3). Select a solution $\textbf{s}' \in \left( N(\textbf{s}) - V \right)$ randomly, and $V= V \cup \left\{ \textbf{s}' \right\}$; Evaluating $\triangle R = R_s^c(\textbf{s}') - R_s^c(\textbf{s})$. \\
       4). If $\triangle R > 0$, then $\textbf{s}=\textbf{s}'$. \\
       5). If $| V | = S$, then go to Step $3.6$; else go to Step $3.3$.  \\
       6). If $R_s^c(\textbf{s}) > R_{pre}^c$, then go to Step $3.2$; else go to Step $4$.
\STATE Annealing process: reducing simulation mutation parameter $C_{k+1} = C_k - \triangle C_k$,
       $\triangle C_k > 0$.
\STATE If 'stop criterion' is not satisfied, i.e., $C_k > C_f$, then setting $k = k+1$, go to Step
       2; otherwise, output: $\textbf{s}_{opt} =\textbf{s}$.
\STATE \textbf{End}
\end{algorithmic}
\end{algorithm}

More explicitly, Algorithm $\ref{Alg2}$ combines the mutation process and self-reproduction strategy into an evolutionary process for approaching the optimum AAG. This search process is performed repeatedly upon the self-reproduction processes and annealing strategy, where the self-reproduction processes search within the immediate neighborhood for an improved solution. Algorithm $\ref{Alg2}$ always starts with a random AAG in reach for a local minimum and then escapes from the suboptimal local 'traps'.
To generate a random neighborhood AAG vector $\textbf{s}' \in N(\textbf{s})$ in Step $2.1$ of Algorithm $2$, the neighborhood sampling technique of Procedure $1$ is proposed. The initial AAG vector $\textbf{s}$ contains $N_s$ '$1$' elements and $N_t-N_s$ '$0$' elements, where the function Randint$[n_1, n_2]$ means that a uniformly distributed integer is randomly generated from the interval $[n_1, n_2]$. Procedure $1$ ensures that the number of elements '$1$' of $\textbf{s}'$ equals to $N_s$.

\floatname{algorithm}{Procedure} \setcounter{algorithm}{0}
\begin{algorithm}
\caption{Generating neighborhood AAG vector}
\begin{algorithmic} [1]
\STATE Given $\textbf{s}= (s_1, \cdots, s_{N_t})$ with $ \sum \nolimits_{i=1}^{N_t} s_i =N_s$.
\STATE Find the indexes with $1$ elements and $0$ elements of $\textbf{s}$, and put them in $\textbf{I}_1 = (e_1, \cdots, e_{N_s})$ and $\textbf{I}_0 = (j_1, \cdots, j_{N_t-N_s})$, respectively.
\STATE Setting $r_1= \textrm{Randint}[1, N_s]$, $r_0= \textrm{Randint}[1, N_t-N_s]$.
\STATE Setting $s_{I_1(r_1)} =0$ and $s_{I_0(r_0)} =1$.
\STATE $\textbf{s}' = (s_1, \cdots, s_{I_1(r_1)}, \cdots, s_{I_0(r_0)},  \cdots, s_{N_t})$.
\RETURN
\end{algorithmic}
\end{algorithm}

As shown in Algorithm $\ref{Alg2}$, the implementation of the proposed SA-based AAG selection scheme requires the design of the following two distinct processes. The first is the process of generating a new solution by Procedure $1$, which exploits a specific generation mechanism and compares the two solutions in term of their cost. Then a decision is made as to whether or not the new AAG could be accepted. The other process is the evolution control strategy, which requires an initial value of the mutation parameter and a decrement function of the mutation parameter.
Remarkably, the annealing process was found to constantly decrease the mutation parameters $C_k$ until it reaches the optimum level.

\subsection{Separate optimization of AAG by SA and of the ANCM by Max-ASR (Separate SA-Max-ASR)}

Similar to \cite{Shu2018two}, the following low complexity method is proposed for  separately optimizing $\textbf{S}_l$ and $\textbf{Q}$. Considering that MI is originated from the active TA indices and APM symbols of the SM system, thus the AAG may be optimized before designing $\textbf{Q}$. Once the AAG has been determined, the corresponding $\textbf{Q}$ is optimized by (\ref{Regular_Q}). According to \cite{Xia2018AS}, the ASR in terms of the AAG is
\begin{align}
R_s^s= I_e^s - I_b^s
\end{align}
where $I_e^s$ is the approximate rate in terms of the AAG for Eve on the face of a realistic channel estimation error, given by
\begin{align}
I_e^s =\textrm{log}_2\sum \limits_{m=1}^{N_sM}\sum \limits_{k=1}^{N_sM}\textrm{exp} \left( \frac{- P_1 \textbf{d}_{mk}^H ( \tilde{\textbf{g}}_l^H  \tilde{\textbf{g}}_l +\sigma_e^2 \textbf{I} ) \textbf{d}_{mk} }{4 \sigma_E^2 } \right).
\end{align}
Similarly, the approximate rate of Bob can be expressed as
\begin{align}
I_b^s = \textrm{log}_2\sum \limits_{i=1}^{N_sM}\sum \limits_{j=1}^{N_sM}\textrm{exp} \left( \frac{- P_1  \textbf{d}_{ij}^H \textbf{h}_l^H  \textbf{h}_l \textbf{d}_{ij} }{4 \sigma_B^2 } \right).
\end{align}
In order to reduce the computational complexity, the repeated calculations can be avoided by defining a pair of upper triangular matrices, $\textbf{U}_B \in \mathbb{R}^{N_t \times N_t}$ and $\textbf{U}_E \in \mathbb{R}^{N_t \times N_t}$, whose $(u,v)$-th entry is respectively given by
\begin{align}
&b_{u,v}= \left\{ \begin{array}{l}
\sum \limits_ {m,n \in \mathbb{M}} \textrm{exp} \left( \frac{- P_1  | h_us_m -h_vs_n |^2 }{4 \sigma_B^2 } \right), u<v \\
\sum  \limits_ {m \neq n \in \mathbb{M}} \textrm{exp} \left( \frac{- P_1  |h_u|^2 | s_m - s_n |^2 }{4 \sigma_B^2 } \right),  u=v, m>n
\end{array} \right.  \\ \nonumber
&e_{u,v}= \\
&\left\{ \begin{array}{l}
\sum \limits_ {m,n \in \mathbb{M}} \textrm{exp} \left( \frac{- P_1  | (|g_u|^2+ \sigma_e^2 )s_m -(|g_v|^2+ \sigma_e^2 )s_n |^2 }{4 \sigma_B^2 } \right), u<v \\
\sum  \limits_ {m \neq n \in \mathbb{M}} \textrm{exp} \left( \frac{- P_1 (|g_u|^2+ \sigma_e^2 ) | s_m - s_n |^2 }{4 \sigma_B^2 } \right),  u=v, m>n
\end{array} \right..
\end{align}
Upon $\textbf{U}_B$ and $\textbf{U}_E$, the AAG associated with the highest SR can be promptly found by utilizing the SA strategy, upon replacing the OF $R_s^c(\textbf{s})$ by
\begin{align}
R_{s}^s= \textrm{log}_2 D_E - \textrm{log}_2 D_B
\end{align}
in Algorithm $\ref{Alg2}$, where $D_E$ and $D_B$ are the summations of sub-triangular entries corresponding to the selected AAG, respectively. Once the AAG is obtained, the corresponding $\textbf{Q}$ is designed by (\ref{Regular_Q}). The separate optimization scheme dramatically reduces the complexity, because the procedure of designing $\textbf{Q}$ is avoided as the AAG changes. Our simulation results will show that this decoupled design strategy strikes a compelling  performance versus complexity tradeoff.

\section{Asymptotic simple equivalence of SR in large-scale scenario and associated optimization}

The proposed joint and separate SA-Max-ASR optimization schemes of Subsections III-B and III-C possess low complexities compared to the direct of Max-SR optimization of Algorithm \ref{Alg1}. However, it may still be a complex task as $N_t$ tends to large values, because the computational complexity of the iterative algorithm grows exponentially upon increasing the number of TAs. With this motivation, we provide a new method that removes the two-layer sum over the legitimate transmit vectors of $\mathcal{S}$, hence dramatically reducing the computational complexity.

\textbf{Theorem 1}: As the number of TAs tends to a large scale, the optimization problem of maximizing the SR of (\ref{Close_ASR}) can be reduced to maximizing the ratio of the $\textrm{SINR}$ at the desired receiver to that at eavesdropper (Max-R-SINR) as follows
\begin{subequations}
\begin{align} \label{Product_Large}
&\max\limits_{\textbf{Q}} ~ R_L'(\textbf{S}_l, \textbf{Q}) \\
&s.t.~~ \textsf{tr}(\textbf{Q}) =1, \textbf{Q} \succeq 0,  \\
&~~~~~~ \sum \limits_{i=1}^{N_t} s_i =N_s,  \\
&~~~~~~ s_i \in \left\{ 0,1 \right\}, i=1, \cdots, N_t,
\end{align}
\end{subequations}
where we have:
\begin{align}
R_L'(\textbf{S}_l, \textbf{Q})=\frac{ \| \textbf{h}_l \|^2 \left( P_2 \tilde{\textbf{g}}_l \textbf{Q} \tilde{\textbf{g}}_l^H + \sigma_E'^2 \right) }{ (\|\tilde{\textbf{g}}_l\|^2 + N_s\sigma_e^2) \left(P_2 \textbf{h}_l\textbf{Q} \textbf{h}_l^H + \sigma_B^2\right)}.
\end{align}
~~~\emph{Proof:} See Appendix A. $\hfill\blacksquare$

Upon comparing $R_L'(\textbf{S}_l, \textbf{Q})$ to (\ref{Close_ASR}), it can be seen that the two-layer summation of legitimate transmit symbols is removed from the OF for large-scale SSM systems. This can be explained as follows: having a large scale provides a high diversity gain, hence the transmit symbols have a mere negligible effect on the received energy of the desired receiver and of the eavesdropper. As a result, the optimization problem can be translated to an energy  maximization  problem that is only related to the communication channels.

Noting that the OF in (\ref{Product_Large}) is a linear fractional function and always non-convex, in accordance with the technique in \cite{Nie2008Trace}, the problem can be rewritten for a given $\mathcal{C}_l$ as
\begin{subequations}  \label{Dink_Large}
\begin{align} \label{Dink_Ob}
&\max\limits_{\textbf{Q}} ~ S_E(\textbf{Q})- \lambda S_B(\textbf{Q}) \\
&s.t.~~~\textbf{Q} \succeq 0  \\
&~~~~~~~ \textsf{tr}(\textbf{Q}) =1.
\end{align}
\end{subequations}
where
\begin{align}
S_E(\textbf{Q})= P_2  \tilde{\textbf{g}}_l \textbf{Q} \tilde{\textbf{g}}_l^H + \sigma_E'^2, \\ S_B(\textbf{Q})= P_2 \textbf{h}_l\textbf{Q} \textbf{h}_l^H + \sigma_B^2,
\end{align}
and $\lambda$ is an auxiliary variable, which is iteratively updated by
\begin{align} \label{Itera}
\lambda [t+1] = \frac{S_E(\textbf{Q}[t])}{S_B(\textbf{Q}[t]},
\end{align}  %\cite{Shen2018}
where $t$ is the iteration index. It has been shown in \cite{Nie2008Trace} that the convergence is guaranteed by alternatively updating $\lambda$ using (\ref{Itera}) and solving it for $\textbf{Q}$ with the aid of (\ref{Dink_Large}), because $\lambda$ is nondecreasing after each iteration.

Upon replacing $R_s^c (\textbf{S}_l, \textbf{Q})$ by $R_L'(\textbf{S}_l, \textbf{Q})$, the SA-based AAG selection can solve the mixed integer optimization problem using Algorithm $\ref{Alg2}$.
Nevertheless, its complexity may still be excessive, because $\textbf{Q}$ is determined by a series of iterations once the AAG changes, and the number of changes is always high. In view of this, it is necessary to design a method that can promptly find a potential AAG to improve the security.
Upon assuming $\textbf{Q}_l^\ast$ is the optimal ANCM for the $l$-th AAG, we have
\begin{align} \nonumber
&\max \limits_{\textbf{S}_l} ~\frac{  \|\textbf{h}_l\|^2  \left( P_2  \tilde{\textbf{g}}_l \textbf{Q}_l^\ast \tilde{\textbf{g}}_l^H + \sigma_E'^2 \right)}{ (  \|\tilde{\textbf{g}}_l\|^2  +\sigma_e^2N_s ) \left( P_2 \textbf{h}_l\textbf{Q}_l^\ast \textbf{h}_l^H + \sigma_B^2 \right)} \\
&~~~~= \max \limits_{\textbf{S}_l} ~\frac{  \|\textbf{h}_l\|^2 }{ (  \|\tilde{\textbf{g}}_l\|^2  +\sigma_e^2N_s )} \frac{ \textsf{tr}\left(  (\tilde{\textbf{g}}_l^H  \tilde{\textbf{g}}_l +  \psi_E\textbf{I}) \textbf{Q}_l^\ast \right) }{\textsf{tr}\left( \textbf{h}_l^H\textbf{h}_l+ \psi_B\textbf{I}) \textbf{Q}_l^\ast \right)}, \label{Large_Covert}
\end{align}
where $\psi_E= \sigma_E'^2/P_2 $ and $\psi_B= \sigma_B^2/P_2 $. According to \cite{Nie2008Trace} \cite{Wang2007Trace}, (\ref{Large_Covert}) is non-convex and hence no closed-form solution exists. However, such a trace ratio problem can be transformed into a simpler ratio tracing problem by sacrificing some of the accuracy. Then, it becomes equivalent to the determinant ratio problem  of \cite{Fukunaga1991},
\begin{align} \nonumber
& \frac{ \|\textbf{h}_l|^2 }{ (  \|\tilde{\textbf{g}}_l\|^2  +\sigma_e^2N_s )} \left( \frac{ \textsf{det} \left(  (\tilde{\textbf{g}}_l^H  \tilde{\textbf{g}}_l +  \psi_E\textbf{I}) \textbf{Q}_l^\ast \right) }{ \textsf{det}\left( \left( \textbf{h}_l^H\textbf{h}_l+ \psi_B\textbf{I} \right) \textbf{Q}_l^\ast \right)}\right) \\ \label{Determinate_Reduce}
&~~= \frac{  \|\textbf{h}_l\|^2 }{ (  \|\tilde{\textbf{g}}_l\|^2  +\sigma_e^2N_s )} \left( \frac{ \textsf{det} \left(  (\tilde{\textbf{g}}_l^H  \tilde{\textbf{g}}_l +  \psi_E\textbf{I}) \right) }{ \textsf{det}\left( \left( \textbf{h}_l^H\textbf{h}_l+ \psi_B\textbf{I} \right) \right)}\right) \\
&~~= \frac{  \| \textbf{h}_l\|^2\left( \psi_E^{N_s}+ \psi_E^{N_s-1}\|\tilde{\textbf{g}}_l\|^2 \right) }{ (\|\tilde{\textbf{g}}_l\|^2  +\sigma_e^2N_s) \left( \psi_B^{N_s}+ \psi_B^{N_s-1}\|\textbf{h}_l\|^2 \right)}.
\end{align}
Equation \eqref{Determinate_Reduce} holds as a result of $\textsf{det}(\textbf{A}\textbf{B})=\textsf{det}(\textbf{A}) \cdot \textsf{det}(\textbf{B})$ when the square matrices $\textbf{A}$ and $\textbf{B}$ have the same size. Thus the optimization problem can be further reduced to
\begin{subequations}
\begin{align}
& \max \limits_{\textbf{S}_l} ~\frac{  \| \textbf{h}_l\|^2\left( \psi_E^{N_s}+ \psi_E^{N_s-1}\|\tilde{\textbf{g}}_l\|^2 \right) }{ (\|\tilde{\textbf{g}}_l\|^2  +\sigma_e^2N_s) \left( \psi_B^{N_s}+ \psi_B^{N_s-1}\|\textbf{h}_l\|^2 \right) }, \\
&~~~~~~ \sum \limits_{i=1}^{N_t} s_i =N_s,  \\
&~~~~~~ s_i \in \left\{ 0,1 \right\}, i=1, \cdots, N_t.
\end{align}
\end{subequations}
In this way, the AAG can be pre-determined before designing $\textbf{Q}$, and the complexity will be dramatically reduced, hence the algorithm can be applied in real-time situations. Once the AAG is determined, the corresponding $\textbf{Q}^\ast$ is obtained by (\ref{Dink_Large}).

\section{Convergence and complexity Analysis}

In this section, we investigate the convergence of the proposed SA-based AAG selection scheme. It is noted that the continuous optimization problem of $\textbf{Q}$ is converted into a concave form for any given AAG, thus $\textbf{Q}$ is optimal during the evolution process of AAG.

\subsection{Proof of Convergence}

As shown in Algorithm $\ref{Alg2}$, the proposed SA-based AAG selection algorithm can be viewed as a stochastic process, where the outcome of each iteration strictly depends on the outcome of the previous iteration. Hence, we model and analyze the SA-based AAG selection method using the theory of finite Markov chains.

In \cite{Tian1996A}, the authors have demonstrated that the Markov chain associated with our SA-based $0$-$1$ programming problem exhibits a strong Markov-like properties, and the condition of asymptotical convergence in the homogeneous case of
\begin{align} \nonumber
& \forall \textbf{s}, \textbf{s}' \in \mathcal{C}, \exists~ p>1, \exists~ \textbf{s}_0, \textbf{s}_1, \cdots, \textbf{s}_p \in \mathcal{C}, \textrm{with} \\
&~~ \textbf{s}_0=\textbf{s}, \textbf{s}_p=\textbf{s}', \textrm{and} ~G_{\textbf{s}_k,\textbf{s}_{k+1}}^c > 0, k=0,\cdots,p-1, \label{M_Prob}
\end{align}
where $G_{\textbf{s}\textbf{s}'}^c$ stands for the generation probability corresponding to the integer sampling procedure of Algorithm $\ref{Alg2}$.

\textbf{Lemma 1.} The generation probability of $G_{\textbf{s},\textbf{s}'}^c$ is equal to
\begin{align} \label{G_Prob}
G_{\textbf{s}, \textbf{s}'}^c = \left\{ \begin{array}{l}
\frac{1}{N_sN_t-N_s^2} ~~\textbf{s}' \in \mathcal{C} \\
~~~~~~0 ~~~~~~  \textbf{s}' \notin \mathcal{C}
\end{array} \right.
\end{align}
with $ \textbf{s}, \textbf{s}' \in \mathcal{C}$.

\textbf{Proof}: A neighborhood AAG vector $\textbf{s}' \in N(\textbf{s})$ of a given AAG vector $\textbf{s} \in \mathcal{C}$ is close to $\textbf{s}$ with
\begin{align} \nonumber
N(\textbf{s}) = &\left\{ \textbf{s}' \in \mathcal{C}:  \textbf{s}'  \textrm{ is constructed by randomly removing one  } \right. \\ \nonumber
&\left. \textrm{TA in}~ \textbf{I}_1 \textrm{ and meanwhile randomly activating } \right. \\ & \left.\textrm{ one silent TA in} ~ \textbf{I}_0 \right\},
\end{align}
where $\textbf{I}_1$ and $\textbf{I}_0$ are the active TA index sets of the current AAG and of the silent TAs, respectively. The integer neighborhood sampling procedure randomly deactivates an active antenna in $\textbf{I}_1$, and randomly activates a silent antenna in $\textbf{I}_0$ at the same time. Therefore the size of neighborhoods is
\begin{align}
|N(\textbf{s})| = N_s(N_t-N_s),
\end{align}
which completes the proof of Lemma $1$. $\hfill\blacksquare$

\textbf{Theorem 2}: The proposed SA-based AAG selection method of maximizing $R_s^c(\textbf{s})$ converges asymptotically to the globally optimal $\textbf{s}_{\textrm{opt}}$, where the generation probabilities $G_{\textbf{s}, \textbf{s}'}^c$ given in (\ref{G_Prob}) asymptotically satisfy the condition in (\ref{M_Prob}) from a global viewpoint.

~~\emph{Proof:} See Appendix B. $\hfill\blacksquare$

For any $\textbf{S}_l$, the optimal solution $\textbf{Q}$ of $R_s^c$ can be obtained as a benefit of its concavity. Upon combining it with the proposed SA-based AAG selection method, we can conclude that our proposed joint Max-ASR scheme can approach the globally optimal $\textbf{S}_l$ and $\textbf{Q}$.

\subsection{Complexity Analysis}

In this subsection, the complexities of the different algorithms are calculated in terms of the number of floating-operations (FLOPs). For the direct solution in Algorithm $\ref{Alg1}$, the computational and search complexity is excessive because a large number of sample points ($N_{\textrm{samp}} \geq 500$) for $\boldsymbol{\Delta}\textbf{g}_l$, $n_B$ and $n_E$ are required to evaluate the accurate SR. The total number of FLOPs of the ES plus GD method of Algorithm $\ref{Alg1}$ can be expressed as
\begin{align}
\mathcal{O}_{\textrm{ES plus GD}}= 3LN_{\textrm{samp}}D_1N_s^2M^2\left( 4N_s^3 + 7N_s^2 + N_s +6 \right),
\end{align}
where $D_1$ is the number of iterations for Algorithm $\ref{Alg1}$. It is remarkable that $L$ will become significantly large as $N_t$ increases.

For our proposed joint SA-Max-ASR in Algorithm $\ref{Alg2}$, the complexity is imposed by three parts: the annealing process, solving the concave maximization problem and the sampling procedure. The complexity of solving problem (\ref{Regular_Q}) each time is \cite{Zhu2017}
\begin{align}
\mathcal{C}_{\textbf{Q}}=& 2M^2N_s^2(3N_s^3 +4N_s^2)  + \mathcal{O}\left[ N_s^{4.5} \textrm{ln}(1/\epsilon)\right],
\end{align}
where the first term denotes the complexity of calculating (\ref{Regular_Expression}).
Therefore, the total complexity of the joint SA-Max-ASR can be expressed as
\begin{align}
\mathcal{O}_{\textrm{Joint SA-Max-ASR}}= KS D_2 \left[ \mathcal{C}_{\textbf{Q}}+ 2M^2N_s^2(2N_s^3 +3N_s^2) \right],
\end{align}
where $D_2$ denotes the number of iterations required for solving (\ref{Regular_Q}), $K$ is the number of mutations necessitated for reaching the termination threshold $C_f$ and $S$ is the size of the samples in Algorithm $\ref{Alg2}$.

For the separate SA-Max-ASR scheme, the pair of matrices $\textbf{D}_B$ and $\textbf{D}_E$ have to be calculated firstly, which requires about $M^2\left( N_t^2 + N_t \right)$ FlOPs. Adding the above complexities and including that of the SA algorithm for finding the AAG, the total complexity of the separate SA-Max-ASR optimization is
\begin{align}
\mathcal{O}_{\textrm{Separate SA-Max-ASR}}=D_2 \mathcal{C}_{\textbf{Q}} + KS M^2\left( N_t^2 + N_t \right).
\end{align}
Once $\textbf{D}_B$ and $\textbf{D}_E$ have been calculated, the SA-based algorithm can be used for rapidly finding the optimal AAGs because only some summation operations are required for the sub-matrices. Moreover, the problem in (\ref{Regular_Q}) only has to be solved once in the separate SA-Max-ASR scheme, thus the computational cost will be dramatically reduced. Based upon the above complexity analysis, it can be observed that the complexity of the proposed joint and separate SA-Max-ASR optimization schemes are much lower than that of Algorithm $\ref{Alg1}$ due to having $N_{\textrm{samp}} \gg N_s$ and $L \gg KS$.

For the large-scale SSM system, the optimization OF is converted into the ratio of $\textrm{SINR}_B$ to $\textrm{SINR}_E$. The complexity of the Max-R-SINR optimization can be approximated as \cite{Chu2016}
\begin{align}
\mathcal{O}_{\textrm{Max-R-SINR}}\approx nD_3 \mathcal{O}(N_s^{3.5}) + 4KSN_s,
\end{align}
where $D_3$ is the number of iterations required for solving (\ref{Dink_Large}).
Compared to the joint SA-Max-ASR, the complexity of the separate SA-Max-ASR optimization reduced to the order of $\mathcal{O}(N_s^{4.5})$.
In large-scale scenarios, our proposed Max-R-SINR optimization method only requires about $\mathcal{O}(N_s^{3.5})$ FLOPs to optimize the variable $\textbf{Q}$, which dramatically reduces the computational complexity.

\section{Simulation Results and Discussions}

In this section, numerical simulation results are presented for evaluating the SR performance of the proposed methods, where the LNSP method \cite{Shu2018two} is used as a performance benchmark. Specifically, the system parameters are set as follows:  $P_s=N_s$ and $C_f=0.001$, the termination condition for all algorithms is set to  $\epsilon=0.0001$ and the initial mutation parameter $C_0$ is computed according to \cite{Ben2004Computing}. The noise levels at the desired receiver and at the eavesdropping receiver are assumed to be identical, i.e., $\sigma_B^2= \sigma_E^2$. Additionally, the ergodic SR is averaged over $500$ channel realizations according to (\ref{Average_SR}). %$P_1=0.25P_s$,

\begin{figure} [ht!]
 \centering
 \includegraphics[width=0.48\textwidth]{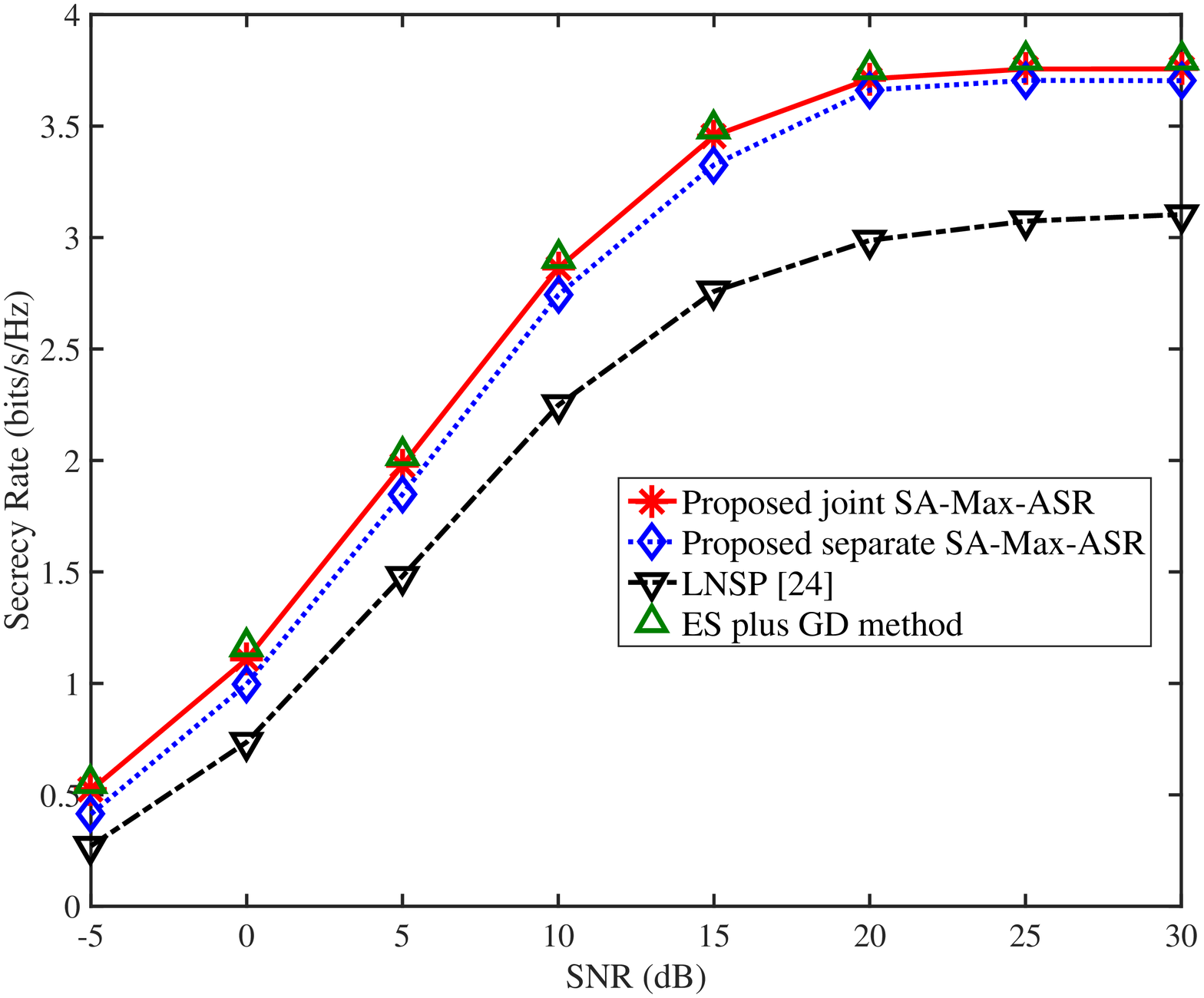}\\
 \caption{ Achievable SR versus SNR with $N_t=7$, $N_s=4$, and $\sigma_e^2=0.25$, and the modulation scheme is QPSK.}\label{SR74}
\end{figure}
Fig. \ref{SR74} plots the achievable SR of the proposed joint and separate SA-Max-ASR optimization methods versus the SNR with $\sigma_e^2=0.25$. It can be seen from this figure that the proposed methods achieve higher SR performance gains than the LNSP method of \cite{Shu2018two}. To be specific, the commonly-used NSP method imposes a serious SR performance loss, because it only considers the interference signal without giving cognizance to the entire secure SM network as a whole. Additionally, the SR performance of the joint SA-Max-ASR optimization is close to that of ES plus GD scheme, in which $5$ random initializations of $\textbf{T}$ are repeated for Algorithm $\ref{Alg1}$.
Remarkably, the separate SA-Max-ASR performs slightly worse than the joint method, while it has a much lower complexity. Hence, it strikes a beneficial performance versus complexity tradeoff.

\begin{figure} [tp!]
 \centering
 \includegraphics[width=0.48\textwidth]{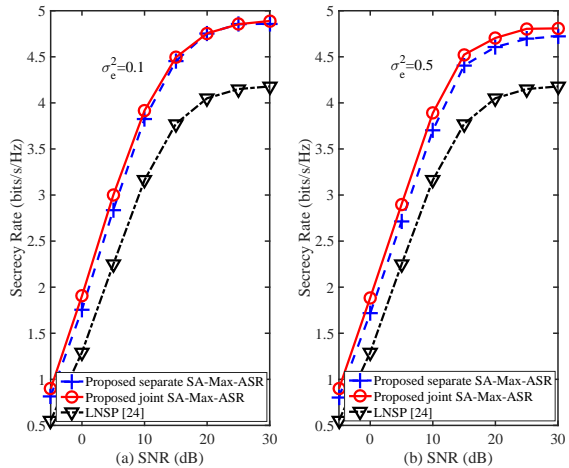}\\
 \caption{Comparison of achievable SR versus SNR in the case of $N_t=15$ and $N_s=8$, where (a) $\sigma_e^2=0.1$ and (b) $\sigma_e^2=0.5$.}\label{ComparisionRobust}
\end{figure}
Fig. \ref{ComparisionRobust} shows the achievable SR of the proposed joint and separate SA-Max-ASR optimization methods for $N_t=15$ and $N_s=8$ with $\sigma_e^2=0.1$ and $0.5$, respectively.
Due to the prohibitive complexity ($L=6435$), we do not consider the performance curve of Algorithm \ref{Alg1} in Fig. \ref{ComparisionRobust} for comparison.
It becomes evident from Fig. \ref{ComparisionRobust} (a) that the SR of our proposed separate SA-Max-ASR optimization is close to that of the joint SA-Max-ASR scheme.
Additionally,  Fig. \ref{ComparisionRobust} (b) shows that the SR performance of our proposed methods is better than that of the LNSP method regardless of the estimation error of the illegitimate channel.
\begin{figure} [tp!]
 \centering
 \includegraphics[width=0.48\textwidth]{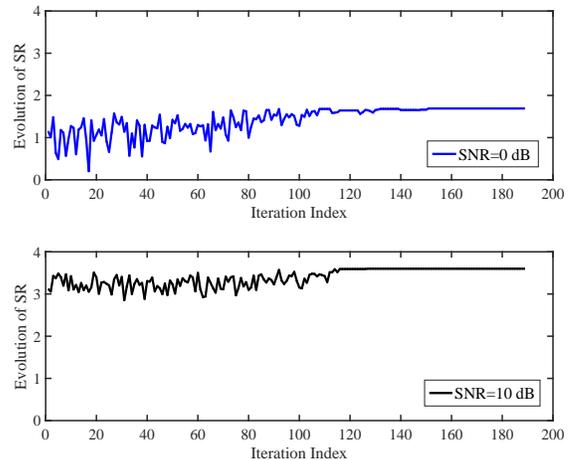}\\
 \caption{Evolution of the achievable SR upon increasing the number of iterations at SNR=0dB and SNR=10dB, using the same configurations as in Fig. \ref{ComparisionRobust} (a). }\label{EvolutionSR}
\end{figure}
Specifically, Fig. \ref{EvolutionSR} shows the evolution process of the achievable SR versus the number of iterations for our proposed joint SA-Max-ASR optimization both at SNR=$0$dB and $10$dB, respectively. Observing the two sub-figures, it follows that the SR performance of the joint SA-Max-ASR optimization eventually converges to a stable level exhibiting slight regional oscillations. In other words, the convergence of the proposed joint SA-Max-ASR scheme can be ensured as a benefit of having a probability of accepting a new generated AAG upon decreasing $C_k$.

Let us now consider the SR performance of the proposed Max-R-SINR optimization scheme.
\begin{figure} [ht!]
 \centering
 \includegraphics[width=0.48\textwidth]{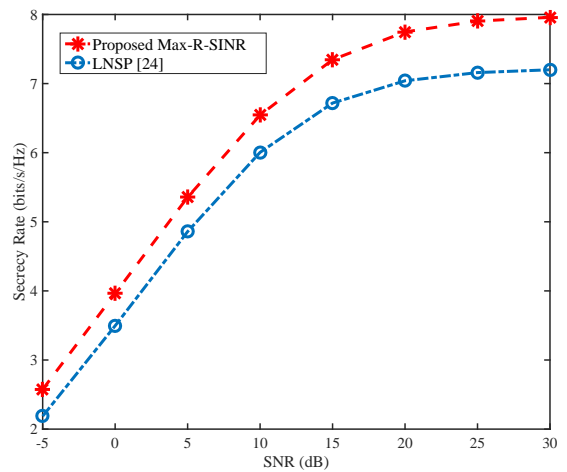}\\
 \caption{ Achievable SR versus SNR with $N_t=100$, $N_s=64$, and $\sigma_e^2=0.25$, and the modulation scheme is QPSK.}\label{SRLarge}
\end{figure}
Fig. \ref{SRLarge} shows the achievable SR versus SNR for $N_t= 100$ and $N_s=64$ with $\sigma_e^2=0.25$. In this case, we let the improvements along the gradients to reach the steepest descent (i.e., $|V|=|N(\textbf{s})|$) in SA. Observe that the proposed Max-R-SINR scheme is capable of providing a significant SR performance benefit over the LNSP scheme. More particularly, the SR performance of our proposed Max-R-SINR scheme approaches $\textrm{log}_2N_sM =8$ in the high-SNR region, which also demonstrates the efficiency of the proposed Max-R-SINR scheme.

\begin{figure} [ht!]
 \centering
 \includegraphics[width=0.48\textwidth]{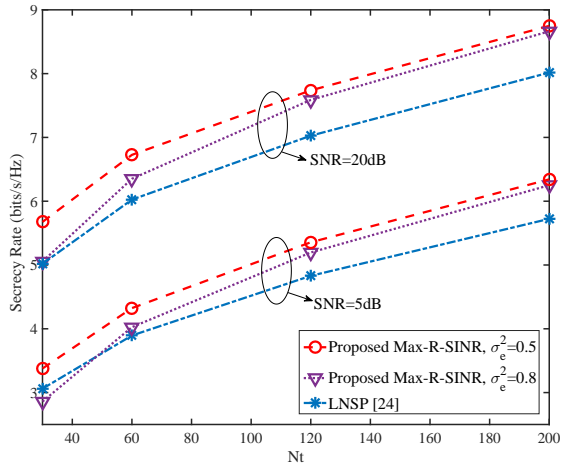}\\
 \caption{ Achievable SR versus the number of TAs for SNR=$5$dB and $20$dB with $\sigma_e^2=0.5$ and $\sigma_e^2=0.8$.}\label{ComparisonTA}
\end{figure}
Finally, in Fig. \ref{ComparisonTA}, we evaluate the accuracy of the proposed Max-R-SINR method for different number of TAs, where $N_t=30, 60, 120, 200$ are considered and the corresponding number of active TAs is $16, 32, 64$ and $128$, respectively.
We observe from Fig. \ref{ComparisonTA} that the SR performance of the proposed Max-R-SINR is low for the LNSP scheme, when $N_t=30$ and $\sigma_e^2=0.8$ at SNR=5dB.
This is because $N_t$ has to reach a certain size to fully exploit the advantage of our proposed Max-R-SINR scheme in terms of its SR. Additionally, it can be seen that the SR performance of the Max-R-SINR becomes better than that of the LNSP for $N_t \geq 60$, and its benefit becomes more apparent upon increasing $N_t$.
Specially, Fig. \ref{ComparisonTA} underlines the merit of Max-R-SINR optimization in terms of its SR when $N_t=200$, albeit $\sigma_e^2$ increases from 0.5 to 0.8.
In summary, we can conclude that the Max-R-SINR scheme efficiently reduces the computational cost in large-scale SSM scenarios.

\newcounter{TempEqCnt} % 创建临时变量TempEqCnt
\setcounter{TempEqCnt}{\value{equation}} % 将当前公式序号 赋给TempEqCnt
\setcounter{equation}{72} % 当前公式序号变为x，x等于长公式应有的序号减1.
\begin{figure*}[ht!]
\begin{align} \nonumber
&\lim _{N_s \rightarrow \infty} R_A (\textbf{S}_l, \textbf{Q}) =\lim _{N_s \rightarrow \infty} \left\{ \textrm{log}_2  \sum \limits_{m=1}^{N_sM}\sum \limits_{k=1}^{N_sM} \textrm{exp} \left( - \frac{1}{4 }  \textbf{d}_{mk}^H \textbf{M}_E \textbf{d}_{mk} \right) - \textrm{log}_2 \sum \limits_{i=1}^{N_sM} \sum \limits_{j=1}^{N_sM}  \textrm{exp}  \left( - \frac{1}{4 }  \textbf{d}_{ij}^H \textbf{M}_B \textbf{d}_{ij} \right)
 \right\}
\\
&~~=\lim _{N_s \rightarrow \infty} \textrm{log}_2 \left\{ \frac{ \sum \limits_{m=1}^{N_sM}\sum \limits_{k=1}^{N_sM} \textrm{exp} \left( - \frac{1}{4 }  \textbf{d}_{mk}^H \textbf{M}_E \textbf{d}_{mk} \right) }
{ \sum \limits_{i=1}^{N_sM} \sum \limits_{j=1}^{N_sM}  \textrm{exp}  \left( - \frac{1}{4 }  \textbf{d}_{ij}^H \textbf{M}_B \textbf{d}_{ij} \right) } \right\}
= \lim _{N_s \rightarrow \infty} \textrm{log}_2 \left\{ \frac{ N_sM+ \sum \limits_{m=1}^{N_sM}\sum \limits_{k=1,k \neq m}^{N_sM} \textrm{exp} \left( - \frac{1}{4 }  \textbf{d}_{mk}^H \textbf{M}_E \textbf{d}_{mk} \right) }
{ N_sM+  \sum \limits_{i=1}^{N_sM} \sum \limits_{j=1,j \neq i}^{N_sM}  \textrm{exp}  \left( - \frac{1}{4 }  \textbf{d}_{ij}^H \textbf{M}_B \textbf{d}_{ij} \right) } \right\} \\
&~~= \lim _{N_s \rightarrow \infty} \textrm{log}_2 \left\{ \frac{ N_sM +  N_s \cdot \sum \limits_{\iota_E=1}^{M(N_sM-1)} \left( \frac{1}{N_s} \sum \limits_{\imath=1}^{N_s} \textrm{exp} \left( - \frac{1}{4 }  \frac{ P_1 \textsf{tr}(\textbf{u} \textbf{u}^H) }{ P_2  \tilde{\textbf{g}}_l \textbf{Q} \tilde{\textbf{g}}_l^H + \sigma_E'^2 } \lambda_{\iota_E} u_\imath^H u_\imath \right) \right) }
{ N_sM + N_s \cdot \sum \limits_{\iota_B=1}^{(N_sM-1)M} \left( \frac{1}{N_s} \sum \limits_{\jmath=1}^{N_s} \textrm{exp}  \left( - \frac{1}{4 }  \frac{ P_1 \textsf{tr}(\textbf{h}_l \textbf{h}_l^H) }{ P_2  \textbf{h}_l \textbf{Q} \textbf{h}_l^H + \sigma_B^2 } \lambda_{\iota_B} u_\jmath^H u_\jmath \right) \right) } \right\} \\
&~~= \textrm{log}_2 \left\{ \frac{ M +  \sum \limits_{\iota_E=1}^{(N_sM-1)M} \left( \int_{u_\imath} \textrm{exp} \left( - \left( \frac{ P_1 \textsf{tr}(\textbf{u} \textbf{u}^H) }{ 4 \left( P_2  \tilde{\textbf{g}}_l \textbf{Q} \tilde{\textbf{g}}_l^H + \sigma_E'^2 \right) }\right)
\lambda_{\iota_E} u_\imath^H  u_\imath \right) d u_\imath \right)}
{ M +  \sum \limits_{\iota_B=1}^{(N_sM-1)M} \left( \int_{u_\jmath} \textrm{exp} \left( - \left( \frac{ P_1 \textsf{tr}(\textbf{h}_l \textbf{h}_l^H) }{ 4 \left( P_2  \textbf{h}_l \textbf{Q} \textbf{h}_l^H + \sigma_B^2 \right) }\right)
\lambda_{\iota_B} u_\jmath^H  u_\jmath \right) d u_\jmath \right) } \right\} \\
&~~ =\textrm{log}_2 \left\{  \frac{M + \sum \limits_{\iota_E=1}^{(N_sM-1)M}  \frac{ 4 \pi\left( P_2  \tilde{\textbf{g}}_l \textbf{Q} \tilde{\textbf{g}}_l^H + \sigma_E'^2 \right)}{ \lambda_{\iota_E} P_1 \textsf{tr}(\textbf{u} \textbf{u}^H)}}{M + \sum \limits_{\iota_B=1}^{(N_sM-1)M} \frac{ 4 \pi\left( P_2  \textbf{h}_l \textbf{Q} \textbf{h}_l^H + \sigma_B^2 \right)}{ \lambda_{\iota_B} P_1 \textsf{tr}(\textbf{h}_l \textbf{h}_l^H)}} \right\}
\approx \textrm{log}_2 \left\{ \frac{ \frac{ \left( P_2  \tilde{\textbf{g}}_l \textbf{Q} \tilde{\textbf{g}}_l^H + \sigma_E'^2 \right)}{ P_1 \textsf{tr}(\textbf{u} \textbf{u}^H)} \left( \frac{1}{\lambda_{1}} + \cdots + \frac{1}{\lambda_{(N_sM-1)M}}\right) }{ \frac{ \left( P_2  \textbf{h}_l \textbf{Q} \textbf{h}_l^H + \sigma_B^2 \right)}{ P_1 \textsf{tr}(\textbf{h}_l \textbf{h}_l^H)} \left( \frac{1}{\lambda_{1}} + \cdots + \frac{1}{\lambda_{(N_sM-1)M}}\right)}  \right\} \\
&~~= \textrm{log}_2 \left\{ \frac{ \frac{ \left( P_2  \tilde{\textbf{g}}_l \textbf{Q} \tilde{\textbf{g}}_l^H + \sigma_E'^2 \right)}{ P_1 \textsf{tr}(\textbf{u} \textbf{u}^H)} }{ \frac{ \left( P_2  \textbf{h}_l \textbf{Q} \textbf{h}_l^H + \sigma_B^2 \right)}{ P_1 \textsf{tr}(\textbf{h}_l \textbf{h}_l^H)} }  \right\}
=\textrm{log}_2 \left\{ \frac{ P_1\| \textbf{h}_l \|^2 \left( P_2 \tilde{\textbf{g}}_l \textbf{Q} \tilde{\textbf{g}}_l^H + \sigma_E'^2 \right) }{ P_1 (\|\tilde{\textbf{g}}_l\|^2 + N_s\sigma_e^2) \left(P_2 \textbf{h}_l\textbf{Q} \textbf{h}_l^H + \sigma_B^2\right)}  \right\}.  \label{NewSR}
\end{align}
\hrulefill
\end{figure*}
\setcounter{equation}{\value{TempEqCnt}}

\section{Conclusions}

In this paper, the joint AAG selection and ANCM design were studied, when only rough partial CSI of Eve is obtained at transmitter. Due to the high-complexity of the ES plus GD method, both a joint and a separate SA-Max-ASR optimization scheme was proposed for optimizing the AAG and ANCM. Compared to the latter, the former achieves a better SR performance which is close to that of the ES plus GD method, while the latter has a lower complexity at the cost of a slight SR performance loss.
To mitigate the complexity of the Max-ASR method when $N_t$ tends to be large, we conceived
the Max-R-SINR scheme. Our simulation results have quantified the SR performance gains of our proposed schemes compared to the existing LNSP method and shown the tradeoff between the SR performance and the complexity.
Our future work will focus on optimizing the AAG selection matrix and ANCM in the face of both desired channel and wiretap channel estimation errors.

\appendices
\section{Proof of Theorem $1$}

Here, we investigate the characteristics of SSM system when $N_t$ tends to large values. For convenience, let us define the matrices
\begin{align} \label{M_E}
\textbf{M}_E= \frac{ P_1 ( \tilde{\textbf{g}}_l^H  \tilde{\textbf{g}}_l +\sigma_e^2 \textbf{I} ) }{ P_2  \tilde{\textbf{g}}_l \textbf{Q} \tilde{\textbf{g}}_l^H + \sigma_E'^2 }, \\
\textbf{M}_B= \frac{ P_1 \textbf{h}_l^H  \textbf{h}_l }{ P_2  \textbf{h}_l \textbf{Q} \textbf{h}_l^H + \sigma_B^2 }.
\end{align}
As $N_t \rightarrow \infty$, the components of the channel vector can be considered to obey the Gaussian distribution. Upon exploiting that
\begin{align}
\textbf{d}_{mk}^H \textbf{M}_E \textbf{d}_{mk} = \left( \frac{ P_1 }{ P_2  \tilde{\textbf{g}}_l \textbf{Q} \tilde{\textbf{g}}_l^H + \sigma_E'^2 }\right)
 \textbf{u}^H \textbf{D}_{mk} \textbf{u},
\end{align}
where $\textbf{D}_{mk}= \textbf{d}_{mk}\textbf{d}_{mk}^H$, and $\tilde{\textbf{g}}_l^H  \tilde{\textbf{g}}_l +\sigma_e^2 \textbf{I} =\textbf{u} \textbf{u}^H$ since $\tilde{\textbf{g}}_l^H  \tilde{\textbf{g}}_l +\sigma_e^2 \textbf{I}$ is symmetric and normalizing $\textbf{u}$ as a standard Gaussian distribution $\bar {\textbf{u}}$, we arrive at
\begin{align} \label{Normal_E}
\left( \frac{ P_1 }{ P_2  \tilde{\textbf{g}}_l \textbf{Q} \tilde{\textbf{g}}_l^H + \sigma_E'^2 }\right)
 \textbf{u}^H \textbf{D}_{mk} \textbf{u} =
\frac{ P_1 \textsf{tr}(\textbf{u} \textbf{u}^H) }{ P_2  \tilde{\textbf{g}}_l \textbf{Q} \tilde{\textbf{g}}_l^H + \sigma_E'^2 } \bar{\textbf{u}}^H \textbf{D}_{mk} \bar {\textbf{u}}.
\end{align}
When $m \neq k$, the rank of $\textbf{D}_{mk}=(\boldsymbol{x}_m-\boldsymbol{x}_k)(\boldsymbol{x}_m-\boldsymbol{x}_k)^H$ equals to 1 and then $\textbf{D}_{mk}$ can be rewritten as $\textbf{D}_{mk}=\textbf{U}_{mk}^H\textsf{diag}(\lambda_{mk},0,\cdots,0)\textbf{U}_{mk}$, where $\lambda_{mk}$ represents the unique nonzero eigenvalue of $\textbf{D}_{mk}$ and $\textbf{U}_{mk}$ is the unitary matrix whose columns are the eigenvectors of $\textbf{D}_{mk}$. As a further step, the expression (\ref{Normal_E}) may be shown to be equivalent to
\begin{align}
\textbf{d}_{mk}^H \textbf{M}_E \textbf{d}_{mk} = \frac{ P_1 \textsf{tr}(\textbf{u} \textbf{u}^H) }{ P_2  \tilde{\textbf{g}}_l \textbf{Q} \tilde{\textbf{g}}_l^H + \sigma_E'^2 } \lambda_{mk} u_\imath^H u_\imath, ~ \left( m \neq k \right),
\end{align}
in which $u_\imath$ is the first element of the vector $\textbf{U}_{mk} \bar{\textbf{u}}$. The components of $\textbf{U}_{mk} \bar{\textbf{u}}$ still follow the Gaussian distribution
due to the fact that the vectors $\bar{\textbf{u}}$ and $\textbf{U}_{mk} \bar{\textbf{u}}$ have the same statistics. When $\lambda_{mk} \neq 0$, as $N_t (N_s) \rightarrow \infty$, we have
\begin{align} \nonumber
&\lim _{N_s \rightarrow \infty}~ \frac{1}{N_s} \sum \limits_{\imath=1}^{N_s} \textrm{exp} \left( -\frac{1}{4} \textbf{d}_{mk}^H \textbf{M}_E \textbf{d}_{mk} \right) \\
&= \int_{u_\imath} \textrm{exp} \left( - \left( \frac{ P_1 \textsf{tr}(\textbf{u} \textbf{u}^H) }{ 4 \left( P_2  \tilde{\textbf{g}}_l \textbf{Q} \tilde{\textbf{g}}_l^H + \sigma_E'^2 \right) }\right)
\lambda_{mk} u_\imath^H  u_\imath \right) d u_\imath, \label{Integer_E}  \\
&= \frac{ 4 \pi\left( P_2  \tilde{\textbf{g}}_l \textbf{Q} \tilde{\textbf{g}}_l^H + \sigma_E'^2 \right)}{ \lambda_{mk} P_1 \textsf{tr}(\textbf{u} \textbf{u}^H)}.
\end{align}
Upon exploiting that there are $N_sM$ possibilities of $\textbf{d}_{mk}^H \textbf{M}_E \textbf{d}_{mk}$ $\left( m, k \in (1, \cdots, N_sM ) \right)$ equal to $0$, we can derive expression (\ref{NewSR}), shown at the top of the previous page, where $\lambda_{ij}$ is the eigenvalue of $\textbf{D}_{ij}=(\boldsymbol{x}_i-\boldsymbol{x}_j)(\boldsymbol{x}_i-\boldsymbol{x}_j)^H$, and $u_\jmath$ is the first component of the vector $\textbf{U}_{ij} \bar{\textbf{u}}$.

It can be inferred from (\ref{NewSR}) that maximizing $R_A(\textbf{S}_l, \textbf{Q})$ in (\ref{Close_ASR}) for our large-scale SSM system can be further reduced to maximizing the ratio of $\textrm{SINR}_B$ at the desired receiver to $\textrm{SINR}_E$ at the eavesdropper, where we have:
\setcounter{equation}{77}
\begin{align}
\textrm{SINR}_B&= \frac{P_1 \textsf{tr}(\textbf{h}_l \textbf{h}_l^H)}{P_2  \textbf{h}_l \textbf{Q} \textbf{h}_l^H + \sigma_B^2},  \\
\textrm{SINR}_E&= \frac{P_1 \textsf{tr} (\tilde{\textbf{g}}_l^H  \tilde{\textbf{g}}_l +\sigma_e^2 \textbf{I})}{P_2  \tilde{\textbf{g}}_l \textbf{Q} \tilde{\textbf{g}}_l^H + \sigma_E'^2 }.
\end{align}
This completes the proof of Theorem $1$.

\section{Proof of Theorem $2$}

It can be seen from Procedure $1$ that two components, $e_i~ (i=1, \cdots, N_s)$ in $\textbf{I}_1$ and $j_c$ $(c=1, \cdots, N_t-N_s)$ in $\textbf{I}_0$ have to be chosen randomly to swap positions, i.e., $e_i \rightarrow 0$ and $j_c \rightarrow 1$. The probability for a neighborhood AAG to be selected equals to $1/(N_sN_t-N_s^2)$, thus we have
\begin{align} \nonumber
& \forall \textbf{s}, \textbf{s}' \in \mathcal{C}, \exists~ p>1, \exists~ \textbf{s}_0, \textbf{s}_1, \cdots, \textbf{s}_p \in \mathcal{C}, \textrm{with} \\
&~~ \textbf{s}_0=\textbf{s}, \textbf{s}_p=\textbf{s}', \textrm{and} ~G_{\textbf{s},\textbf{s}'}^c > 0, k=0,\cdots,p-1.
\end{align}
Consequently, the components of the stationary distribution $\textbf{q}(C)$ of the Markov chain $\xi(C)$ satisfy
\begin{align} \nonumber
q_\textbf{s}(C)& = \lim \limits_ {k \rightarrow \infty} P \left\{ \xi_C(k)=\textbf{s}|\xi_C(0)=\textbf{s}' \right\}  \\
&=\frac{|N(\textbf{s})|\textrm{exp} \left( \frac{-R_s^c(\textbf{s}) }{C}\right) }{\sum \limits_{\textbf{s}' \in \mathcal{C}} |N(\textbf{s}')| \textrm{exp} \left( \frac{-R_s^c(\textbf{s}') }{C} \right) }.
\end{align}
Then,
\begin{align}
\lim \limits_ {C \rightarrow 0} q_\textbf{s}(C) = q_\textbf{s}^\ast = \left\{ \begin{array}{l}
\frac{1}{|S_{\textrm{opt}}|} ~~\textbf{s} \in S_{\textrm{opt}} \\
~~0 ~~~~  \textbf{s} \notin S_{\textrm{opt}}.
\end{array} \right.
\end{align}
Finally, we have
\begin{align}
\lim \limits_{C \rightarrow 0 } \lim \limits_{k \rightarrow \infty } P \left\{ \xi_C(k)=\textbf{s} \right\} = \lim \limits_{C \rightarrow 0} q_\textbf{s}(C) =q_\textbf{s}^\ast ,
\end{align}
or
\begin{align}
\lim \limits_{C \rightarrow 0 } \lim \limits_{k \rightarrow \infty } P \left\{ \xi_C(k) \in S_{\textrm{opt}} \right\} = \sum \limits_{s \in S_{\textrm{opt}}} q_\textbf{s}^\ast =1,
\end{align}
which completes the proof of Theorem $2$.

\bibliographystyle{IEEEtran}
\bibliography{IEEEabrv,refer}

\begin{thebibliography}{10}
\providecommand{\url}[1]{#1}
\csname url@samestyle\endcsname
\providecommand{\newblock}{\relax}
\providecommand{\bibinfo}[2]{#2}
\providecommand{\BIBentrySTDinterwordspacing}{\spaceskip=0pt\relax}
\providecommand{\BIBentryALTinterwordstretchfactor}{4}
\providecommand{\BIBentryALTinterwordspacing}{\spaceskip=\fontdimen2\font plus
\BIBentryALTinterwordstretchfactor\fontdimen3\font minus
  \fontdimen4\font\relax}
\providecommand{\BIBforeignlanguage}[2]{{%
\expandafter\ifx\csname l@#1\endcsname\relax
\typeout{** WARNING: IEEEtran.bst: No hyphenation pattern has been}%
\typeout{** loaded for the language `#1'. Using the pattern for}%
\typeout{** the default language instead.}%
\else
\language=\csname l@#1\endcsname
\fi
#2}}
\providecommand{\BIBdecl}{\relax}
\BIBdecl

\bibitem{Mesleh2008Spatial}
R.~Y. Mesleh, H.~Haas, S.~Sinanovic, C.~W. Ahn, and S.~Yun, ``Spatial
  modulation,'' \emph{IEEE Trans. on Veh. Technol.}, vol.~57, no.~4, pp.
  2228--2241, Jul. 2008.

\bibitem{Basar2017}
E.~Basar, M.~Wen, R.~Mesleh, M.~D. Renzo, Y.~Xiao, and H.~Haas, ``Index
  modulation techniques for next-generation wireless networks,'' \emph{IEEE
  Access}, vol.~5, pp. 16\,693--16\,746, 2017.

\bibitem{YangP2013}
P.~{Yang}, B.~{Zhang}, Y.~{Xiao}, B.~{Dong}, S.~{Li}, M.~{El-Hajjar}, and
  L.~{Hanzo}, ``Detect-and-forward relaying aided cooperative spatial
  modulation for wireless networks,'' \emph{IEEE Trans. on Commun.}, vol.~61,
  no.~11, pp. 4500--4511, Nov. 2013.

\bibitem{Basar2016}
E.~Basar, ``On multiple-input multiple-output {OFDM} with index modulation for
  next generation wireless networks,'' \emph{IEEE Trans. on Signal Process.},
  vol.~64, no.~15, pp. 3868--3878, Aug. 2016.

\bibitem{HeL2017}
L.~{He}, J.~{Wang}, J.~{Song}, and L.~{Hanzo}, ``On the multi-user multi-cell
  massive spatial modulation uplink: How many antennas for each user?''
  \emph{IEEE Trans. on Wireless Commun.}, vol.~16, no.~3, pp. 1437--1451, Mar.
  2017.

\bibitem{Renzo2013Spatial}
M.~D. Renzo, H.~Haas, A.~Ghrayeb, S.~Sugiura, and L.~Hanzo, ``Spatial
  modulation for generalized {MIMO}: Challenges, opportunities, and
  implementation,'' \emph{Proc. IEEE}, vol. 102, no.~1, pp. 56--103, Jan. 2014.

\bibitem{YangP2016Precod}
P.~{Yang}, Y.~L. {Guan}, Y.~{Xiao}, M.~D. {Renzo}, S.~{Li}, and L.~{Hanzo},
  ``Transmit precoded spatial modulation: Maximizing the minimum euclidean
  distance versus minimizing the bit error ratio,'' \emph{IEEE Trans. on
  Wireless Commun.}, vol.~15, no.~3, pp. 2054--2068, Mar. 2016.

\bibitem{Lakshmi2015}
T.~{Lakshmi Narasimhan}, P.~{Raviteja}, and A.~{Chockalingam}, ``Generalized
  spatial modulation in large-scale multiuser {MIMO} systems,'' \emph{IEEE
  Trans. on Wireless Commun.}, vol.~14, no.~7, pp. 3764--3779, Jul. 2015.

\bibitem{WuQ2017}
Q.~Wu, G.~Y. Li, W.~Chen, D.~W.~K. Ng, and R.~Schober, ``An overview of
  sustainable green 5{G} networks,'' \emph{IEEE Wireless Commun.}, vol.~24,
  no.~4, pp. 72--80, Aug. 2017.

\bibitem{Jin2017}
J.~Jin, C.~Xiao, M.~Tao, and W.~Chen, ``Linear precoding for fading cognitive
  multiple-access wiretap channel with finite-alphabet inputs,'' \emph{IEEE
  Trans. on Veh. Technol.}, vol.~66, no.~4, pp. 3059--3070, Apr. 2017.

\bibitem{ChenX2017}
X.~Chen, D.~W.~K. Ng, W.~H. Gerstacker, and H.~H. Chen, ``A survey on
  multiple-antenna techniques for physical layer security,'' \emph{IEEE Commun.
  Surv. Tut.}, vol.~19, no.~2, pp. 1027--1053, 2017.

\bibitem{ZouY2016}
Y.~{Zou}, J.~{Zhu}, X.~{Wang}, and L.~{Hanzo}, ``A survey on wireless security:
  Technical challenges, recent advances, and future trends,'' \emph{Proc.
  IEEE}, vol. 104, no.~9, pp. 1727--1765, Sep. 2016.

\bibitem{WangF2018}
F.~{Wang}, C.~{Liu}, Q.~{Wang}, J.~{Zhang}, R.~{Zhang}, L.~{Yang}, and
  L.~{Hanzo}, ``Optical jamming enhances the secrecy performance of the
  generalized space-shift-keying-aided visible-light downlink,'' \emph{IEEE
  Trans. on Commun.}, vol.~66, no.~9, pp. 4087--4102, Sep. 2018.

\bibitem{Wang2012Distributed}
H.~M. Wang, Q.~Yin, and X.~G. Xia, ``Distributed beamforming for physical-layer
  security of two-way relay networks,'' \emph{IEEE Trans. on Signal Process.},
  vol.~60, no.~7, pp. 3532--3545, Jul. 2012.

\bibitem{Wu2012Linear}
Y.~Wu, C.~Xiao, Z.~Ding, X.~Gao, and S.~Jin, ``Linear precoding for
  finite-alphabet signaling over {MIMOME} wiretap channels,'' \emph{IEEE Trans.
  on Veh. Technol.}, vol.~61, no.~6, pp. 2599--2612, Jul. 2012.

\bibitem{Wu2017Secure}
Y.~Wu, J.~B. Wang, J.~Wang, R.~Schober, and C.~Xiao, ``Secure transmission with
  large numbers of antennas and finite alphabet inputs,'' \emph{IEEE Trans. on
  Commun.}, vol.~65, no.~8, pp. 3614--3628, Aug. 2017.

\bibitem{Liu2017Secure}
C.~Liu, L.~L. Yang, and W.~Wang, ``Secure spatial modulation with a full-duplex
  receiver,'' \emph{IEEE Wireless Commun. Lett.}, vol.~6, no.~6, pp. 838--841,
  Dec. 2017.

\bibitem{Wang2015Secrecy}
L.~Wang, S.~Bashar, Y.~Wei, and R.~Li, ``Secrecy enhancement analysis against
  unknown eavesdropping in spatial modulation,'' \emph{IEEE Commun. Lett.},
  vol.~19, no.~8, pp. 1351--1354, Aug. 2015.

\bibitem{Chen2016Secure}
Y.~Chen, L.~Wang, Z.~Zhao, M.~Ma, and B.~Jiao, ``Secure multiuser {MIMO}
  downlink transmission via precoding-aided spatial modulation,'' \emph{IEEE
  Commun. Lett.}, vol.~20, no.~6, pp. 1116--1119, Jun. 2016.

\bibitem{WuF2016}
F.~Wu, R.~Zhang, L.~Yang, and W.~Wang, ``Transmitter precoding-aided spatial
  modulation for secrecy communications,'' \emph{IEEE Trans. on Veh. Technol.},
  vol.~65, no.~1, pp. 467--471, Jan. 2016.

\bibitem{Wu2015Secret}
F.~Wu, L.~L. Yang, W.~Wang, and Z.~Kong, ``Secret precoding-aided spatial
  modulation,'' \emph{IEEE Commun. Lett.}, vol.~19, no.~9, pp. 1544--1547, Sep.
  2015.

\bibitem{Jiang2018}
X.~Q. Jiang, M.~Wen, H.~Hai, J.~Li, and S.~Kim, ``Secrecy-enhancing scheme for
  spatial modulation,'' \emph{IEEE Commun. Lett.}, vol.~22, no.~3, pp.
  550--553, Mar. 2018.

\bibitem{YangY2018}
Y.~Yang and M.~Guizani, ``Mapping-varied spatial modulation for physical layer
  security: Transmission strategy and secrecy rate,'' \emph{IEEE J. Sel. Area.
  Commun.}, vol.~36, no.~4, pp. 877--889, Apr. 2018.

\bibitem{Shu2018two}
F.~Shu, Z.~Wang, R.~Chen, Y.~Wu, and J.~Wang, ``Two high-performance schemes of
  transmit antenna selection for secure spatial modulation,'' \emph{IEEE Trans.
  on Veh. Technol.}, vol.~67, no.~9, pp. 8969--8973, Sep. 2018.

\bibitem{Xia2018AS}
G.~Xia, F.~Shu, Y.~Zhang, J.~Wang, S.~ten Brink, and J.~Speidel, ``Antenna
  selection method of maximizing secrecy rate for green secure spatial
  modulation,'' \emph{IEEE Trans. on Green commun. and networking, Online},
  DOI: 10.1109/TGCN.2019.2898442.

\bibitem{Rajashekar2013Antenna}
R.~Rajashekar, K.~V.~S. Hari, and L.~Hanzo, ``Antenna selection in spatial
  modulation systems,'' \emph{IEEE Commun. Lett.}, vol.~17, no.~3, pp.
  521--524, Mar. 2013.

\bibitem{Zhou2014Reduced}
Z.~Zhou, N.~Ge, and X.~Lin, ``Reduced-complexity antenna selection schemes in
  spatial modulation,'' \emph{IEEE Commun. Lett.}, vol.~18, no.~1, pp. 14--17,
  Jan. 2014.

\bibitem{Sun2017Transmit}
Z.~Sun, Y.~Xiao, P.~Yang, S.~Li, and W.~Xiang, ``Transmit antenna selection
  schemes for spatial modulation systems: Search complexity reduction and
  large-scale {MIMO} applications,'' \emph{IEEE Trans. on Veh. Technol.},
  vol.~66, no.~9, pp. 8010--8021, Sep. 2017.

\bibitem{YangP2016}
P.~{Yang}, Y.~{Xiao}, Y.~L. {Guan}, S.~{Li}, and L.~{Hanzo}, ``Transmit antenna
  selection for multiple-input multiple-output spatial modulation systems,''
  \emph{IEEE Trans. on Commun.}, vol.~64, no.~5, pp. 2035--2048, May 2016.

\bibitem{WangBo2018}
B.~Wang, H.~Xie, X.~Xia, and X.~Zhang, ``A {NSGA-II} algorithm hybridizing
  local simulated-annealing operators for a bicriteria robust job-shop
  scheduling problem under scenarios,'' \emph{IEEE Trans. on Fuzzy Syst.},
  Early Access, 2018.

\bibitem{Zhan2018}
S.~Zhan, Z.~Zhang, L.~Wang, and Y.~Zhong, ``List-based simulated annealing
  algorithm with hybrid greedy repair and optimization operator for 0¨c1
  knapsack problem,'' \emph{IEEE Access}, vol.~6, pp. 54\,447--54\,458, 2018.

\bibitem{Tang2013}
Y.~Tang, J.~Xiong, D.~Ma, and X.~Zhang, ``Robust artificial noise aided
  transmit design for {MISO} wiretap channels with channel uncertainty,''
  \emph{IEEE Commun. Lett.}, vol.~17, no.~11, pp. 2096--2099, Nov. 2013.

\bibitem{Yu2018}
X.~Yu, Y.~Hu, Q.~Pan, X.~Dang, N.~Li, and M.~H. Shan, ``Secrecy performance
  analysis of artificial-noise-aided spatial modulation in the presence of
  imperfect {CSI},'' \emph{IEEE Access}, vol.~6, pp. 41\,060--41\,067, 2018.

\bibitem{Aghdam2017Joint}
S.~R. Aghdam and T.~M. Duman, ``Joint precoder and artificial noise design for
  {MIMO} wiretap channels with finite-alphabet inputs based on the cut-off
  rate,'' \emph{IEEE Trans. on Wireless Commun.}, vol.~16, no.~6, pp.
  3913--3923, Jun. 2017.

\bibitem{Zhu2017}
Z.~Zhu, Z.~Chu, N.~Wang, S.~Huang, Z.~Wang, and I.~Lee, ``Beamforming and power
  splitting designs for {AN}-aided secure multi-user {MIMO} {SWIPT} systems,''
  \emph{IEEE Trans. on Inf. Forensics Security}, vol.~12, no.~12, pp.
  2861--2874, Dec. 2017.

\bibitem{Nie2008Trace}
F.~Nie, S.~Xiang, Y.~Jia, C.~Zhang, and S.~Yan, ``Trace ratio criterion for
  feature selection,'' in \emph{National Conference on Artificial
  Intelligence}, 2008, pp. 671--676.

\bibitem{Wang2007Trace}
H.~Wang, S.~Yan, D.~Xu, X.~Tang, and T.~Huang, ``Trace ratio vs. ratio trace
  for dimensionality reduction,'' in \emph{IEEE Conference on Computer Vision
  and Pattern Recognition}, 2007, pp. 1--8.

\bibitem{Fukunaga1991}
K.~Fukunaga, ``Introduction to statistical pattern recognition,''
  \emph{Academic Press, second ed.}, 1991.

\bibitem{Tian1996A}
P.~Tian, H.~Wang, and D.~Zhang, ``A darwin and boltzmann mixed strategy for
  global optimization,'' \emph{Journal of Shanghai Jiaotong University}, 1996.

\bibitem{Chu2016}
Z.~Chu, Z.~Zhu, M.~Johnston, and S.~Y.~L. Goff, ``Simultaneous wireless
  information power transfer for {MISO} secrecy channel,'' \emph{IEEE Trans. on
  Veh. Technol.}, vol.~65, no.~9, pp. 6913--6925, Sep. 2016.

\bibitem{Ben2004Computing}
W.~Ben-Ameur, ``Computing the initial temperature of simulated annealing,''
  \emph{Computational Optimization and Applications}, vol.~29, no.~3, pp.
  369--385, 2004.

\end{thebibliography}

\ifCLASSOPTIONcaptionsoff
  \newpage
\fi

\end{document}